\documentclass[a4paper,11pt]{article}

\usepackage{jinstpub}
\usepackage{graphicx}
\usepackage{hyperref}
\usepackage{siunitx}
\usepackage{upgreek}
\usepackage[titletoc]{appendix}
\usepackage{multirow}
\usepackage{longtable}

\usepackage{pifont}
\usepackage{color}

\usepackage[intoc,refpage,noprefix]{nomencl}
\makenomenclature
\renewcommand{\nomname}{List of abbreviations}

\newcommand{\fref}[1]{figure~\ref{#1}}
\newcommand{\figref}[1]{Figure~\ref{#1}}
\newcommand{\tref}[1]{table~\ref{#1}}

\newcommand{\eref}[1]{eq.~(\ref{#1})}
 
\newcommand{\sref}[1]{section~\ref{#1}}
\newcommand{\secref}[1]{Section~\ref{#1}}

\def\kr{\textsuperscript{83m}Kr}
\def\t2{$\text{T}_2$}
\def\betael{$\upbeta$-electron}
\def\betaels{$\upbeta$-electrons}
\def\betadec{$\upbeta$-decay}
\def\betaspec{$\upbeta$-spectrum}
\def\h2o{$\text{H}_2\text{O}$}
\def\ke{KATRIN experiment}
\def\pin{\textit{p-i-n}}

\DeclareSIUnit\year{yr}

\providecommand{\keywords}[1]{\textbf{\textit{JINST Keywords---}} #1}

\title{First transmission of electrons and ions through the KATRIN beamline}

\affiliation[a]{Helmholtz-Institut f\"{u}r Strahlen- und Kernphysik, Rheinische Friedrich-Wilhelms Universit\"{a}t Bonn, Nussallee 14-16, 53115 Bonn, Germany}
\affiliation[b]{Institute of Experimental Particle Physics~(ETP), Karlsruhe Institute of Technology~(KIT), Wolfgang-Gaede-Str. 1, 76131 Karlsruhe, Germany}
\affiliation[c]{Institut f\"{u}r Physik, Johannes-Gutenberg-Universit\"{a}t Mainz, 55099 Mainz, Germany}
\affiliation[d]{Institute for Data Processing and Electronics~(IPE), Karlsruhe Institute of Technology~(KIT), Hermann-von-Helmholtz-Platz 1, 76344 Eggenstein-Leopoldshafen, Germany}
\affiliation[e]{Institute for Nuclear Research of Russian Academy of Sciences, 60th October Anniversary Prospect 7a, 117312 Moscow, Russia}
\affiliation[f]{Institute for Technical Physics~(ITeP), Karlsruhe Institute of Technology~(KIT), Hermann-von-Helmholtz-Platz 1, 76344 Eggenstein-Leopoldshafen, Germany}
\affiliation[g]{Max-Planck-Institut f\"{u}r Kernphysik, Saupfercheckweg 1, 69117 Heidelberg, Germany}
\affiliation[h]{Max-Planck-Institut f\"{u}r Physik, F\"{o}hringer Ring 6, 80805 M\"{u}nchen, Germany}
\affiliation[i]{Technische Universit\"{a}t M\"{u}nchen, James-Franck-Str. 1, 85748 Garching, Germany}
\affiliation[j]{Institute for Nuclear Physics~(IKP), Karlsruhe Institute of Technology~(KIT), Hermann-von-Helmholtz-Platz 1, 76344 Eggenstein-Leopoldshafen, Germany}
\affiliation[k]{Laboratory for Nuclear Science, Massachusetts Institute of Technology, 77 Massachusetts Ave, Cambridge, MA 02139, USA}
\affiliation[l]{Center for Experimental Nuclear Physics and Astrophysics, and Dept.~of Physics, University of Washington, Seattle, WA 98195, USA}
\affiliation[m]{Nuclear Physics Institute of the CAS, v. v. i., CZ-250 68 \v{R}e\v{z}, Czech Republic}
\affiliation[n]{Institut f\"{u}r Kernphysik, Westf\"{a}lische Wilhelms-Universit\"{a}t M\"{u}nster, Wilhelm-Klemm-Str. 9, 48149 M\"{u}nster, Germany}
\affiliation[o]{Department of Physics, Faculty of Mathematics and Natural Sciences, University of Wuppertal, Gauss-Str. 20, D-42119 Wuppertal, Germany}
\affiliation[p]{Department of Physics, Carnegie Mellon University, Pittsburgh, PA 15213, USA}
\affiliation[q]{Universidad Complutense de Madrid, Instituto Pluridisciplinar, Paseo Juan XXIII, n\textsuperscript{\b{o}} 1, 28040 - Madrid, Spain}
\affiliation[r]{Department of Physics and Astronomy, University of North Carolina, Chapel Hill, NC 27599, USA}
\affiliation[s]{Triangle Universities Nuclear Laboratory, Durham, NC 27708, USA}
\affiliation[t]{Commissariat \`{a} l'Energie Atomique et aux Energies Alternatives, Centre de Saclay, DRF/IRFU, 91191 Gif-sur-Yvette, France}
\affiliation[u]{University of Applied Sciences~(HFD)~Fulda, Leipziger Str.~123, 36037 Fulda, Germany}
\affiliation[v]{Department of Physics, Case Western Reserve University, Cleveland, OH 44106, USA}
\affiliation[w]{Institute for Nuclear and Particle Astrophysics and Nuclear Science Division, Lawrence Berkeley National Laboratory, Berkeley, CA 94720, USA}
\affiliation[x]{Institut f\"{u}r Physik, Humboldt-Universit\"{a}t zu Berlin, Newtonstr. 15, 12489 Berlin, Germany}
\affiliation[y]{Project, Process, and Quality Management~(PPQ), Karlsruhe Institute of Technology~(KIT), Hermann-von-Helmholtz-Platz 1, 76344 Eggenstein-Leopoldshafen, Germany    }

\author[a]{M.~Arenz,}
\author[b]{W.-J.~Baek,}
\author[c]{M.~Beck,}
\author[d]{A.~Beglarian,}
\author[b]{J.~Behrens,}
\author[d]{T.~Bergmann,}
\author[e]{A.~Berlev,}
\author[f]{U.~Besserer,}
\author[g]{K.~Blaum,}
\author[h,i]{T.~Bode,}
% ...

\collaboration{KATRIN collaboration}

\emailAdd{dparno@cmu.edu}
\emailAdd{hickford@uni-wuppertal.de}
\emailAdd{alexander.jansen@kit.edu}
\emailAdd{kathrin.valerius@kit.edu}

\date{\today}

\abstract{
The Karlsruhe Tritium Neutrino (KATRIN) experiment is a large-scale effort to probe the absolute neutrino mass scale with a sensitivity of \SI{0.2}{\electronvolt} (90\% confidence level), via a precise measurement of the endpoint spectrum of tritium \betadec{}. 
This work documents several KATRIN commissioning milestones: the complete assembly of the experimental beamline, the successful transmission of electrons from three sources through the beamline to the primary detector, and tests of ion transport and retention. In the First~Light commissioning campaign of Autumn~2016, photo\-electrons were generated at the rear wall and ions were created by a dedicated ion source attached to the rear section; in July~2017, gaseous \kr{} was injected into the KATRIN source section, and a condensed \kr{} source was deployed in the transport section. In this paper we describe the technical details of the apparatus and the configuration for each measurement, and give first results on source and system performance. We have successfully achieved transmission from all four sources, established system stability, and characterized many aspects of the apparatus.
}


\begin{document}
\maketitle
\flushbottom

\section{Introduction}
\label{sec:intro}

Twenty years ago, the definitive observation of neutrino flavor oscillation~\cite{Fukuda:SuperK1998,Ahmad:SNO2002} established that there are three distinct neutrino mass states $\nu_1, \nu_2, \nu_3$, each a coherent superposition of the neutrino flavor states $\nu_\mathrm{e}, \nu_{\mu}, \nu_{\tau}$. 
Since then, oscillation experiments have shown that at least two neutrino-mass states have non-zero mass. The existence of neutrino mass is the first observed contradiction of the Standard Model, and the mass scale is of great interest in fields from particle theory~\cite{de-gouvea:2016} to cosmology~\cite{abazajian:2016}.

In the quasi-degenerate regime, where the mass splittings between neutrino states are small compared to the mass scale ($m_1 \approx m_2 \approx m_3$), the effective mass-squared of the electron anti-neutrino may be defined according to:

\begin{equation}
	m_{\bar{\nu}_\mathrm{e},\, \mathrm{eff}}^2 = \sum^3_{i=1} |U_{\mathrm{e}i}|^2 m_i^2,
\end{equation}

\noindent where $U_{\mathrm{e}i}$ are the elements of the neutrino-mixing matrix connecting the electron neutrino flavor to the $i$th neutrino mass state~\cite{pdg17}. A small, but non-zero value of $m_{\bar{\nu}_\mathrm{e},\, \mathrm{eff}}^2$ imprints a shift and shape distortion on the energy spectrum in \betadec{}s. The tiny signature can best be probed by precision spectroscopy close to the kinematic endpoint.

Following in a series of neutrino-mass experiments~\cite{Robertson1988, Otten2008}, including the Mainz~\cite{Kraus2005} and Troitsk~\cite{TroitskFinal2003, Aseev2011} experiments that set the current laboratory limit of $m_{\bar{\nu}_\mathrm{e},\, \mathrm{eff}} < $ \SI{2}{\electronvolt} (\SI{95}{\percent} confidence level), the Karlsruhe Tritium Neutrino (KATRIN\nomenclature{KATRIN}{Karlsruhe Tritium Neutrino experiment}) experiment will soon start to probe the neutrino-mass scale through this nearly model-independent kinematic method.  
Like its predecessors, KATRIN will exploit the low $Q$-value (\SI{18.6}{\kilo\electronvolt}) and favorable half-life (\SI{12.3}{\year}) of tritium \betadec{}:

\begin{equation}
	^3\mathrm{H} \rightarrow {^3\mathrm{He}^+} + \mathrm{e}^- + \bar{\nu}_\mathrm{e}.
\end{equation}

KATRIN is designed to achieve a neutrino-mass sensitivity of \SI{0.2}{\electronvolt} at the \SI{90}{\percent} confidence level through a high-precision measurement of the integrated endpoint spectrum of tritium \betadec{}~\cite{KATRIN2005}. Since the experimental observable is the square of the neutrino mass, a ten-fold improvement in mass sensitivity requires reducing statistical and systematic uncertainties by a factor of \num{100}. The design, commissioning and calibration requirements on the experiment are accordingly severe.

Details of the apparatus are found in \sref{sec:apparatus}. Briefly, gaseous \t2{} molecules decay within a strong magnetic field in a windowless source section. The resulting \betaels{} are adiabatically guided along magnetic field lines through an extensive transport section, which reduces tritium flow, before entering a pair of integrating electromagnetic spectrometers for energy analysis. Low-energy \betaels{} are reflected back towards the source; high-energy \betaels{} are transmitted to a silicon detector on the far side. The kinetic-energy threshold set by the voltage of the second, high-resolution spectrometer is systematically scanned so as to measure the shape of the integrated spectrum. Calibration and monitoring tools, including supplementary detectors and a third spectrometer for high-voltage monitoring, are implemented throughout the experiment.

In this work, we describe a suite of commissioning measurements performed with the entire KATRIN beamline, demonstrating for the first time that all KATRIN subsystems function together as a complete experiment. During these measurements, we tested four calibration sources essential for KATRIN commissioning and running.

In the First Light campaign (\sref{sec:FirstLight-campaign}), conducted in autumn~2016, we demonstrated the use of photo\-electrons as an electron source (\sref{sec:FirstLightFirstElectrons}), a proof of principle for a planned space-charge neutralization strategy during neutrino-mass running. The photo\-electron source and a second, pencil-beam electron source were used for detailed alignment and transport studies (\sref{sec:FirstLightAlignment}), as well as for the first measurement of KATRIN backgrounds induced by a large electron flux (\sref{sec:FirstLightBackground}).  

In the second part of the First Light measurement phase, we studied ion transport through the experiment (\sref{sec:ion_detection}). During neutrino-mass running, T$^+$ ions and tritiated cluster ions, created as byproducts of tritium \betadec{} and \betael{} ionization, are a potential background source. Our non-radioactive ion source allowed critical tests of experimental strategies for extracting these ions from the beamline (\sref{sec:ionblocking}, \sref{sec.unbeabsichtigteIonenblockierung}). 

In July~2017, we commissioned KATRIN with \kr{} conversion-electron sources of two types. Neutrino-mass analysis requires a precise understanding of \betael{} transport through the magnetic flux tube that connects the source section with the detector. The metastable isotope \kr{}, generated through the decay of \textsuperscript{83}Rb, is an ideal calibration source for these studies, since the K-line energy is near the tritium \betadec{} $Q$-value and the short \kr{} half-life minimizes the risk of contamination. 

In the gaseous \kr{} campaign (\sref{sec:GKrS-campaign}), we introduced \kr{} gas to the source section and used the main spectrometer to scan the conversion-electron lines. This allowed us to study an isotropic source of electrons with well-defined energies, spatially distributed throughout the source section. It also provided a baseline measurement of a calibration source that will later be used to probe local electric potentials in the presence of tritium. We successfully demonstrated all three beamline electron detectors for the first time (\sref{subsec:GKrS:first-electrons}), tested our understanding of the distribution of radioactive gas within our system (\sref{subsec:GKrS:activity-as-seen-in-fpd-and-kr-distribution}), measured line shapes (\sref{subsec:GKrS:qualitative-information-about-line-shapes}) and established system stability (\sref{subsec:GKrS:system-stability}).

During neutrino-mass running, however, the source section will be too cold to permit gaseous \kr{} flow at the desired pressures~\cite{leming:1970}. A second, condensed \kr{} source, which can be inserted into the beamline without a time-consuming system warmup, has therefore been installed. We successfully demonstrated the use of this source for the first time in the condensed \kr{} campaign (\sref{sec:CKrS-campaign}), immediately following the measurements with gaseous \kr{}. The condensed \kr{} source illuminates a small region of the detector (\sref{ssec:ckrs_first_electrons}) with a high enough intensity to permit even the line scans of relatively weak conversion lines (\sref{ssec:ckrs_qualitative_information_about_line_shapes}), and its stability (\sref{ssec:ckrs_stability}) has now been established. 

In these three measurement campaigns, we have established and studied electron and ion transport through the entire KATRIN beamline; commissioned and characterized both sources and detectors; and tested our data-taking and analysis methods in preparation for neutrino-mass measurements with tritium.
\section{Apparatus overview}
\label{sec:apparatus}
\begin{figure}[tbp]
	\centering
	\includegraphics[width=\textwidth]{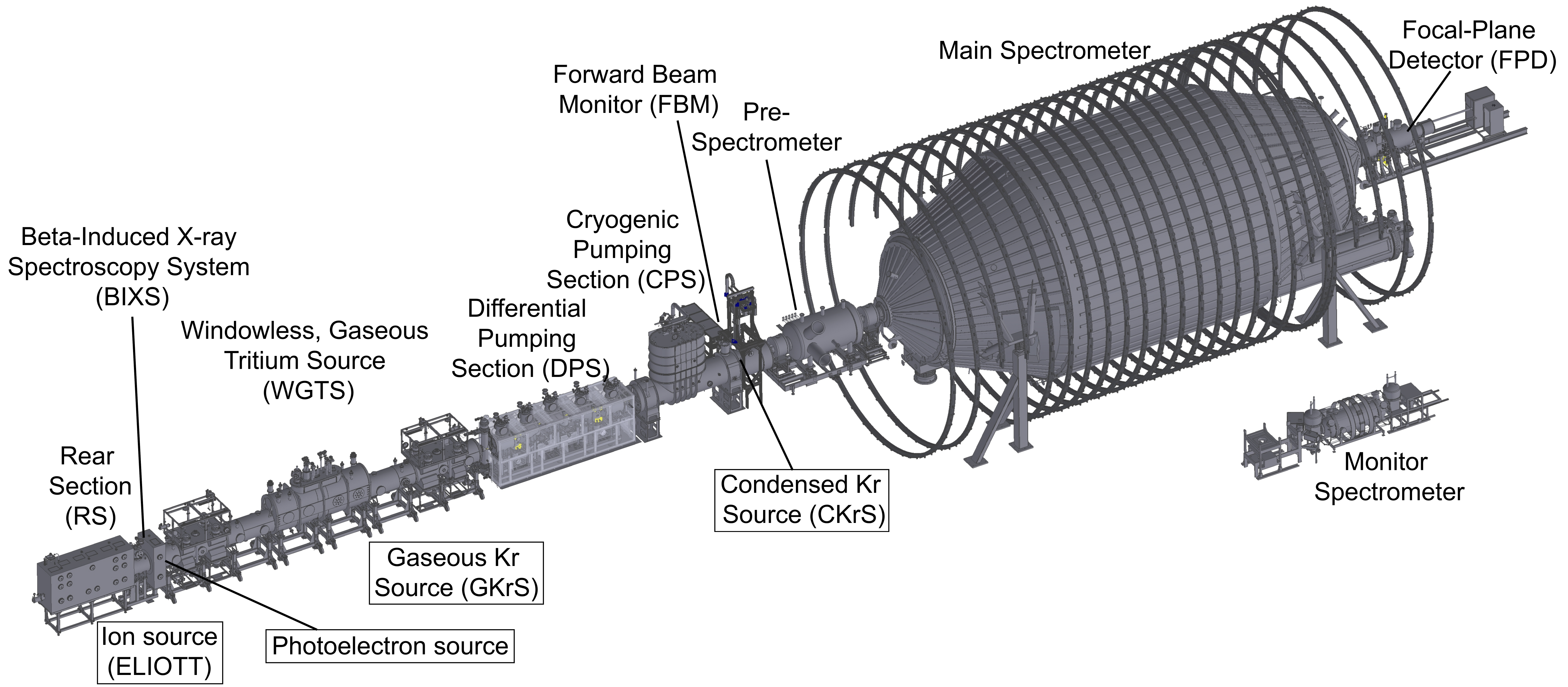}
	\caption{Engineering drawing of the KATRIN apparatus. Calibration sources used in the beamline are labeled in boxes.}
	\label{fig:beamline}
\end{figure}

The design specifications of the KATRIN experiment --- including sub-eV electron energy resolution, tritium decay rates at \SI{e11}{\becquerel}, minimal \betael{} energy loss, and high stability --- necessitate a highly complex suite of equipment, outlined in \sref{sec:beamline} and depicted in overview in \fref{fig:beamline}. In \sref{sec:sources}, we describe in detail the electron and ion sources used for these commissioning measurements. \secref{sec:calmonitor} describes the high-voltage system, the auxiliary detectors for calibration, and the various monitoring systems. 

\subsection{KATRIN beamline}
\label{sec:beamline}
\figref{fig:beamline} gives an overview of the major components of the KATRIN experiment. Starting at the left-hand side of the figure and moving to the right in the downstream direction, the \SI{70}{\metre} KATRIN beamline comprises the following key elements. A rear section (\sref{sec:rearsection}) for calibration and monitoring is connected to the windowless, gaseous source (\sref{sec:wgts}) into which \t2{} gas will be injected during neutrino-mass measurements. The source is followed by two transport sections (\sref{apparatus:transport}) --- the differential and cryogenic pumping sections --- to remove tritium while adiabatically guiding \betaels{} toward the paired integrating spectrometers (\sref{apparatus:sds}). Those \betaels{} that pass the energy threshold of the main spectrometer are counted in the focal-plane detector system (\sref{sec:fpd}). The data-acquisition system is described in \sref{sec:daq}.  

\subsubsection{Rear section}
\label{sec:rearsection}
On the rear or upstream side, the KATRIN beamline is terminated by the rear wall: a gold-plated, stainless-steel disk. Through a central aperture of \SI{5}{\milli\metre} diameter, an electron or ion beam can be sent through the whole beamline. A dedicated electron gun, which was not yet installed for the measurements described in this paper, will produce electrons of well-defined energy and pitch angle with respect to the magnetic field, according to the working principle described in ref.~\cite{Behrens2017}. The narrow electron beam can be steered into any portion of the flux tube for calibration purposes. 

Tritium \betadec{} in the source (\sref{sec:wgts}) naturally produces positive ions, both as decay remnants and as ionization products from the strong \betael{} flux, creating a low-density plasma. All \betael{}s follow the magnetic guiding field either toward the detector or onto the rear wall, where they are absorbed. The resulting positive plasma potential would generate an ion flux toward the rear wall, compensating for this charge loss, but such a potential would distort the spectrum and must be prevented. For this purpose, high-intensity ultraviolet (UV\nomenclature{UV}{ultraviolet}) radiation, similar to the irradiation source described in \sref{sec:rw_pe}, will be used to stimulate emission of an experimentally optimized flux of photo\-electrons from the gold-plated rear wall.
The photo\-electrons also increase the charge density, maintaining the quasi-neutrality of the plasma, which then follows the rear-wall potential via an ambipolar flux.

\subsubsection{Windowless, gaseous source section}
\label{sec:wgts}
The Windowless Gaseous Tritium Source (WGTS\nomenclature{WGTS}{Windowless Gaseous Tritium Source}) is the entry point of tritium into the KATRIN experiment. Hydrogen gas can be prepared at defined concentrations and isotopic composition~\cite{Sturm2010,Schloesser2013c}, and is then introduced into the \SI{10}{\metre}-long WGTS beam tube via the inner loop system~\cite{Priester2015}, a dedicated injection, circulation and purification system, and circulated through the WGTS to maintain the stable gas column acting as tritium source. The decay electrons and ions are guided towards the differential pumping section (\sref{apparatus:transport}) and rear section (\sref{sec:rearsection}) by superconducting solenoids around the beam tube~\cite{Gil2018}. The \SI{10}{\metre}-long beam tube connects at each end to two pumping chambers interleaved with two \SI{1}{\metre}-long vacuum pipes. Neutral gas, e.g.\ \t2{}, is evacuated by a total of \num{14} turbomolecular pumps (TMPs\nomenclature{TMP}{Turbomolecular pump}); the pumped gas then flows into the inner loop system for reprocessing. The TMPs form a first stage of differential pumping, reducing the downstream gas flow by a factor of \num{100}. For calibration purposes, radioactive \kr{} (\sref{sec:GKrS}) can be injected into the last pumping chamber on the forward side of the source cryostat. 

In order to achieve temperature stabilization on the required \SI{0.1}{\percent} level~\cite{KATRIN2005}, each of the five beam-tube segments is equipped with a two-phase cooling system~\cite{Grohmann2009,Grohmann2011,Grohmann2013}. Neon is used during standard operation at \SI{30}{\kelvin} for tritium data-taking and argon in \kr{} mode at \SI{100}{\kelvin}. The central temperature sensors can be calibrated absolutely in a region between \SI{28}{\kelvin} and \SI{32}{\kelvin} by means of a dedicated vapor pressure system. The achieved temperature stability was measured during the gaseous \kr{} commissioning campaign (\sref{subsubsec:GKrS:system-stability:slow-control-stability}).

\subsubsection{Transport section}
\label{apparatus:transport}
The transport section, consisting of the Differential Pumping Section (DPS\nomenclature{DPS}{Differential Pumping Section}) and the Cryogenic Pumping Section (CPS\nomenclature{CPS}{Cryogenic Pumping Section}), has two primary functions. First, it adiabatically guides \betaels{} from the exit of the WGTS cryostat (\sref{sec:wgts}) to the entrance of the spectrometer section (\sref{apparatus:sds}). Second, it reduces the tritium gas flow by some twelve orders of magnitude in order to prevent tritium contamination of the main spectrometer, which would elevate backgrounds. In the following we will describe the characteristics of the two pumping stages.

\begin{figure}[tbp]
	\centering
	\includegraphics[width=\linewidth]{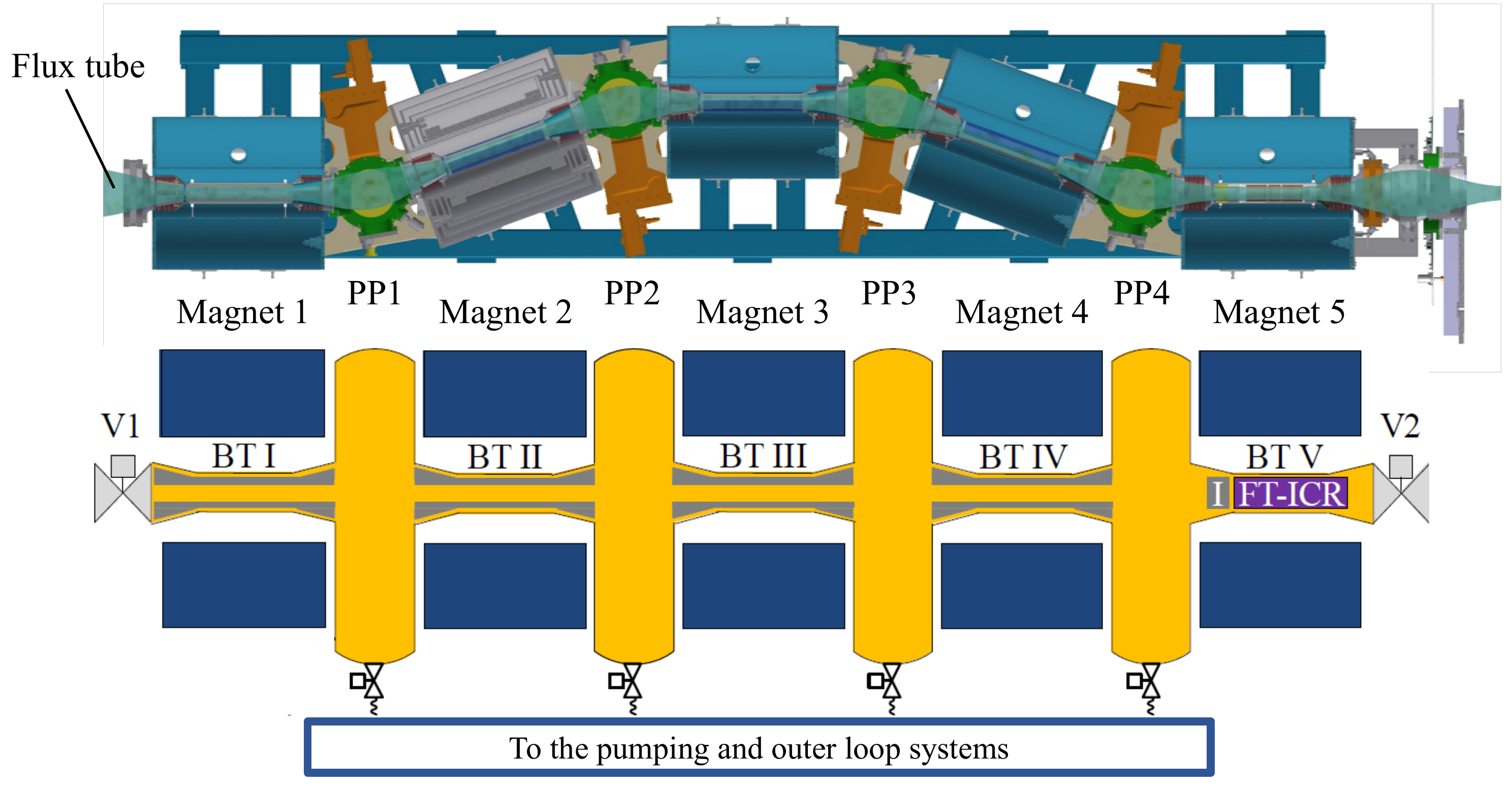}
	\caption{Layout of the Differential Pumping Section (DPS). Top: The engineering model, seen from the top, showing the envelope of the flux tube. Bottom: A schematic diagram, seen from the side, indicating the electrodes. The five distinct beam-tube elements (BT~I to BT~V) are located within the warm bores of the solenoid magnets; the four pump ports, which are connected to the pumping system, are in between the magnets. Each of the elements BT~I--BT~IV hosts a dipole electrode (gray in bottom panel). BT~V hosts a ring electrode (I) and the Fourier transform-ion cyclotron resonance (FT-ICR) device.}
	\label{fig:dps}
\end{figure}

The DPS is designed to achieve a gas-flow reduction by at least five orders of magnitude through a series of turbomolecular pumps~\cite{PhDKosmider2012,PhDJansen2015}.  
Magnetic guiding for \betaels{} is provided by five superconducting solenoids in independent cryostats (\fref{fig:dps}). The DPS beam line consists of five beam-tube elements (BT\nomenclature{BT}{Beam Tube element of DPS}~I to BT~V) placed inside the warm bores of the solenoids, connected by four pump ports between the magnets. Each pump port is connected with a gate valve to a TMP, which pumps the tritium gas via a second TMP stage into the outer loop of the closed KATRIN tritium cycle~\cite{KATRIN2005, PhDSturm2010}. Neighboring beam-tube elements of the DPS have a relative horizontal tilt of \SI{20}{\degree} to prevent direct forward beaming of neutral tritium-containing molecules through the DPS~\cite{Lukic2012}, and to enhance the pumping efficiency. Tritium-containing ions might still be guided by the magnetic field through the DPS beamline, however. To avoid this, a ring electrode is placed inside the entrance of BT~V to block ions, and electric dipole elements --- each consisting of two half-shells with conical ends --- are installed inside BT~I through BT~IV to drift ions out from the flux tube. As a diagnostic tool, a Fourier transform-ion cyclotron resonance (FT-ICR\nomenclature{FT-ICR}{Fourier transform-ion cyclotron resonance}) device is placed inside BT~V downstream of the ring electrode~\cite{UbietoDiaz2009}. The dipole electrode in BT~I was installed following the results of ion tests during the First Light measurement campaign (\sref{sec.unbeabsichtigteIonenblockierung}). The FT-ICR in BT~V was not yet operational during any of the measurements described in this paper.

The CPS~\cite{Gil2010}, the downstream stage of the transport section, must reduce the tritium flow by at least seven orders of magnitude between its entry and its exit. To achieve this high tritium retention, the CPS beam tube acts as an approximately \SI{3}{\metre}-long cold trap. Seven superconducting solenoids guide \betaels{} adiabatically through the CPS; neutral tritium molecules strike the beam-tube wall due to the tilt of the second and fourth solenoid magnets by \SI{15}{\degree} against the spectrometer axis. There, molecules are adsorbed by a \SI{3}{\kelvin} argon-frost layer. The argon-frost layer provides a higher binding energy than normal cryo-condensation, as well as a larger surface area ($A\approx$~\SI{2.57}{\metre\squared}).  To achieve these low temperatures, the cryostat has a nitrogen (\SI{77}{\kelvin}) and a helium loop (\SI{4.5}{\kelvin}, \SI{1.29}{\bar}). The final \SI{3}{\kelvin} cooling of the cold trap is realized by pumping down the liquid helium bath of a heat exchanger to \SI{250}{\milli\bar}. The cold trap will be regenerated after \SI{60}{days} of operation, corresponding to a stored tritium activity of \SI{1}{Ci} at nominal source intensities. A ring electrode for ion blocking, similar to the one in BT~V of the DPS, is positioned at the entrance to the CPS.

\subsubsection{Integrating spectrometers}
\label{apparatus:sds}
The two tandem KATRIN spectrometers --- the pre-spectrometer and main spectrometer --- are of the MAC-E type, operating according to the principle of magnetic adiabatic collimation with electrostatic filtering\nomenclature{MAC-E}{Magnetic adiabatic collimation with electrostatic filtering}~\cite{Beamson1980}. This spectrometer technique has been successfully used to measure integrated tritium beta spectra in the earlier Mainz~\cite{Kraus2005} and Troitsk~\cite{Aseev2011} neutrino-mass experiments. Below, we briefly describe the working principle of the MAC-E filter.

Electrons are created in a magnetic field and guided adiabatically along magnetic field lines to the spectrometer entrance. In this motion, the electron orbital magnetic moment $\mu$ is constant as the electrons travel from a region of strong magnetic field at the spectrometer entrance to a region of weak magnetic field at the central analyzing plane. The electron momenta are thus rotated to be nearly longitudinal, producing a broad, collimated beam of electrons whose kinetic energy can be analyzed by a longitudinal retarding potential. A MAC-E spectrometer can be approximated as an integrating high-pass filter: electrons with energy below the threshold are reflected upstream, while electrons surpassing the energy threshold are re-accelerated on the downstream side of the analyzing plane and recorded by the focal-plane detector system.

\begin{figure}[bpt]
	\centering
	\includegraphics[width=.7\linewidth]{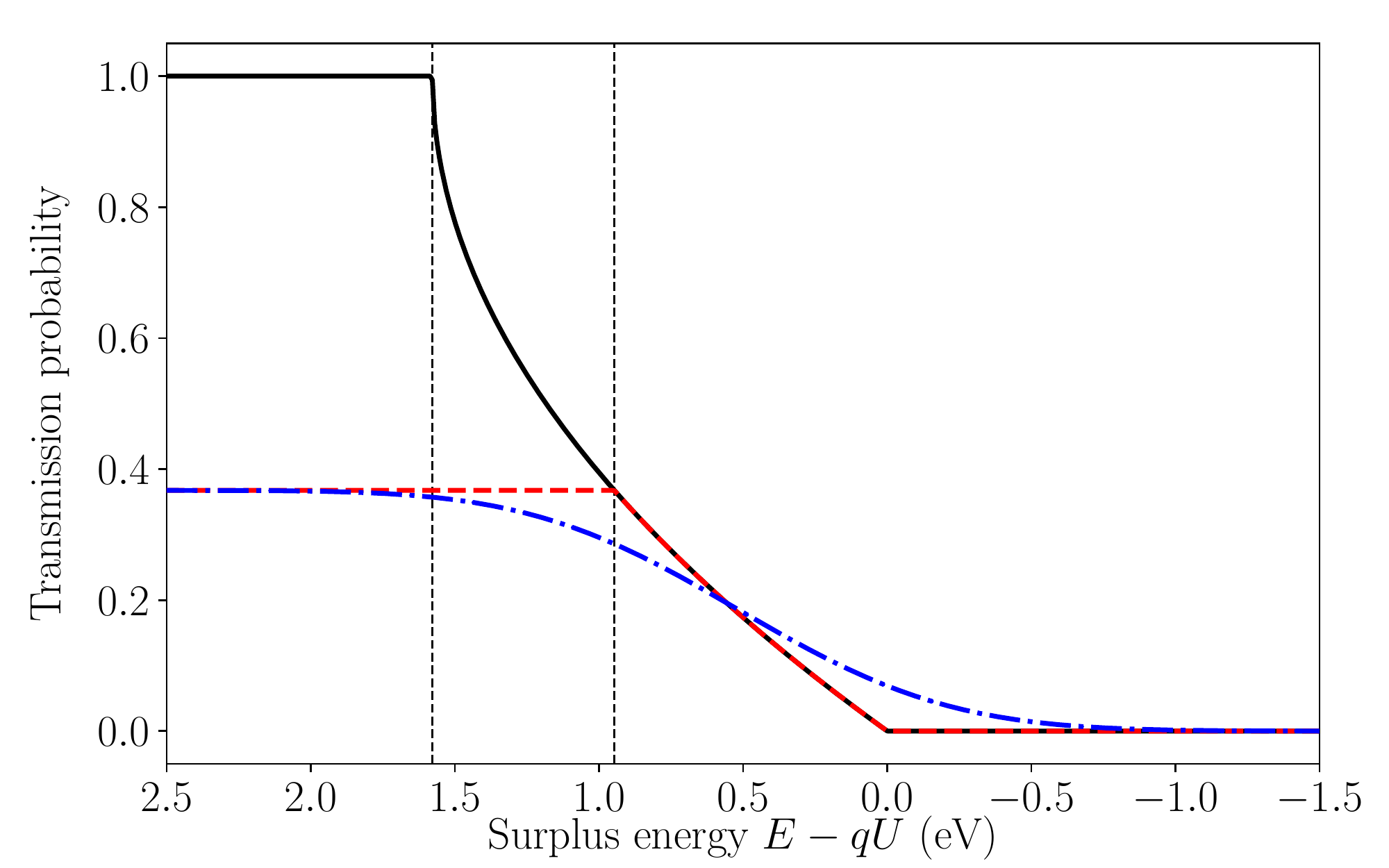}
	\caption{Transmission function of a MAC-E filter (eq. 2.1) for electrons from a mono-energetic, isotropic source with energy $E = $ 18.6 keV; note that the filter energy $qU$ of the spectrometer increases towards the right. In a setup with $B_{\mathrm{max}} = B_{\mathrm{S}} =$ \SI{3.6}{\tesla} and $B_{\mathrm{min}} =$ \SI{0.3}{\milli\tesla}, the energy resolution is $\Delta E =$ \SI{1.58}{\electronvolt} (solid black line). For KATRIN, the maximum magnetic field is reached at the main-spectrometer exit with $B_{\mathrm{max}} =$ \SI{6.0}{\tesla} $> B_{\mathrm{S}}$, and the energy resolution improves to $\Delta E =$ \SI{0.95}{\electronvolt} (dashed red line). For comparison, the dash-dotted blue line shows a transmission function for an isotropic, non-mono-energetic source, featuring a Gaussian energy distribution with a \SI{0.5}{\electronvolt} standard deviation.}
	\label{fig:transmission}
\end{figure}

The probability for an electron with kinetic energy $E$ and charge $q$ to pass the spectrometer, with its filter energy $qU$ defined by the retarding potential $U$, can be described by the transmission function:

\begin{equation}
\label{eq:transmission}
T(E, qU) = \left\{\begin{array}{ll}
    0
        &\quad E - qU < 0 \\
    1 - \displaystyle\sqrt{ 1 - \frac{E - qU}{E} \frac{2}{\gamma + 1} \frac{B_{\mathrm{max}}}{B_{\mathrm{min}}} }
        &\quad 0 \le E - qU \le \Delta E \\
    1 - \displaystyle\sqrt{ 1 - \frac{B_\mathrm{S}}{B_{\mathrm{max}}} }
        &\quad E - qU > \Delta E
\end{array}\right.,
\end{equation}
\noindent where $B_{\mathrm{min}}$ and $B_{\mathrm{max}}$ are the minimum and maximum magnetic fields. For signal electrons at \SI{18.6}{keV}, the relativistic gamma factor $\gamma = \frac{E}{m_\text{e}} + 1 \approx \num{1.04}$ can often be neglected in a non-relativistic approximation.
The transmission probability rises from zero to its maximal amplitude over an energy range that corresponds to $\Delta E$, the filter width or energy resolution of the spectrometer (\fref{fig:transmission}). It is defined by the ratio of $B_{\mathrm{min}}$ to $B_{\mathrm{max}}$:

\begin{equation}
\label{eq:mac-e}
    \frac{\Delta E}{E} = \frac{B_{\mathrm{min}}}{B_{\mathrm{max}}} \frac{\gamma + 1}{2}.
\end{equation}

In the case of KATRIN, the magnetic-field strength in the source, $B_\mathrm{S}$, is lower than the maximum magnetic field, located in the detector section (\sref{sec:fpd}). This leads to an effect known as magnetic reflection, which prevents electrons with large pitch angles relative to the magnetic field from being transmitted, and results in a reduced amplitude of the transmission function. This feature of the electromagnetic setup improves the energy resolution and reduces certain systematic effects~\cite{KATRIN2005}.

The retarding potential, provided by the precision high-voltage system described in \sref{par:high-voltage}, exhibits an inevitable radial variation on the order of \SI{1}{\volt} across the analyzing plane of the MAC-E filter: the actual potential is slightly more positive near the spectrometer axis, compared with the values at larger radii, closer to the electrodes. This potential depression is taken into account in the analysis.

The pre-spectrometer~\cite{Prall2012} is located between the CPS and the main spectrometer. As designed, it will be operated during neutrino-mass measurements at a voltage a few hundred volts below the endpoint energy of tritium \betadec{}. In this way a large fraction of the \betaels{} will be reflected before they reach the main spectrometer, where they could induce background via scattering (\sref{sec:FirstLightBackground}). At the same time, the pre-spectrometer accelerates positive ions out of the flux tube and prevents them from reaching the main spectrometer. For additional ion blocking, a ring electrode, identical to the one in the DPS (\sref{apparatus:transport}), is located in each of the superconducting magnets on either side of the pre-spectrometer. These ring electrodes are redundant to the ion-blocking in the DPS and block a small flux of non-tritium ions, which are created in the CPS via ionization of residual gas by \betaels{}. Additionally, the pre-spectrometer provides additional pumping for any residual neutral tritium.

When the pre-spectrometer is operated at high potential without having been vacuum-baked, the elevated pressure of $\sim$~\SI{e-9}{\milli\bar} allows strong Penning discharges in the region between the two spectrometers~\cite{PhDValerius2009,PhDFraenkle2010,PhDHillen2011,PhDPrall2011}. The pre-spectrometer was therefore kept at ground potential for most of the measurements presented here. 

The KATRIN main spectrometer, located between the pre-spectrometer and the detector section, is the largest and most advanced MAC-E-type spectrometer ever built. It performs energy analysis of \betaels{} with high luminosity as well as unprecedented precision and stability.

The main spectrometer is a \SI{23.2}{\meter} long stainless-steel vacuum vessel of \SI{9.8}{\meter} diameter, equipped with superconducting solenoids at the entrance and exit to create the magnetic guiding field. It is operated at a pressure of a few times \SI{e-11}{\milli\bar}~\cite{Arenz2016}. A normal-conduction air-coil system~\cite{Glueck2013,Erhard2017} around the vessel is used to compensate for the Earth's magnetic field and for the fringe fields of the nearby solenoids, and to correct and fine-tune the low-field region of the filter. The desired filter width, and thus the energy resolution, can be controlled according to \eref{eq:mac-e} via adjustment of these air-coils relative to the superconducting solenoids.

In order to provide the electrostatic retarding potential, the vacuum vessel itself can be maintained at a voltage of up to $U =$~\SI{-35}{\kilo\volt}. A two-layer inner-electrode system, consisting of segmented rings made out of wire modules, is used to fine-tune the retarding potential. By charging the inner electrode about \SI{100}{\volt} more negative than the vessel wall, low-energy secondary electrons created on the vessel wall can be repelled from the flux tube.

Since the electrostatic filter potential is of primary importance for precise and stable electron spectroscopy, the direct-current (DC\nomenclature{DC}{Direct current}) component of the high voltage is monitored by a precision voltage divider on the parts-per-million (ppm\nomenclature{ppm}{parts per million}) level~\cite{Thuemmler2009}. In addition, the high voltage is directly connected to the monitor spectrometer (\sref{sec:mos}), where mono-energetic conversion electrons from an implanted \textsuperscript{83}Rb/\kr{} source are observed as an energy reference~\cite{Erhard2014}. In \sref{par:high-voltage} the high-voltage subsystem is described in detail.

The electromagnetic configuration of the main spectrometer gives rise to several processes producing background electrons in the region of interest, which have been studied extensively in prior commissioning work. The dominant background arises from the ionization of highly excited atoms in the volume~\cite{Fraenkle2017}. Based on previous commissioning measurements, an average background rate of 0.5 counts per second (cps\nomenclature{cps}{counts per second}) is expected in the detector region of interest when imaging the full spectrometer volume.

\subsubsection{Focal-plane detector system}
\label{sec:fpd}
The Focal-Plane Detector (FPD\nomenclature{FPD}{Focal Plane Detector}) system~\cite{Amsbaugh2015} is the final component in the KATRIN beamline for \betaels{} in the energy range of interest. The superconducting pinch magnet completes the MAC-E filter of the main spectrometer, and the electrons are further accelerated using a post-acceleration electrode (PAE\nomenclature{PAE}{Post-acceleration electrode}, at \SI{10}{\kilo\volt}) towards a segmented electron detector. 

A second superconducting magnet ensures that the flux tube fits onto the surface of the detector. A calibration system, housed in the space between the two magnets, includes an $^{241}$Am $\upgamma$ source and a Ti disk to serve as a photoelectron source. The disk may also be used as a beamline Faraday cup, and is equipped with the Precision Ultra-Low Current Integrating Normalization Electrometer for Low-Level Analysis (PULCINELLA\nomenclature{PULCINELLA}{Precision Ultra-Low Current Integrating Normalization Electrometer for Low-Level Analysis})~\cite{martin:phd2017,Amsbaugh2015} to provide a current measurement.

The electron detector is a multi-pixel monolithic silicon \pin{}\nomenclature[p]{\pin{}}{\textit{positive-intrinsic-negative}} diode array on a single wafer with a sensitive area of \SI{65.3}{\centi\metre\squared}. This wafer is segmented into one central bullseye region surrounded by \num{12} concentric rings; the bullseye is quartered, and each ring is divided into \num{12} segments as can be seen in \fref{pic.firstlight}. The \num{148} detector pixels have equal surface area. This segmentation allows radial and azimuthal variations in the experimental response function to be incorporated in the analysis. The pre-amplification stage of the front-end electronics sits directly behind the flange on which the detector is mounted, in a separate vacuum system at a higher operational pressure. Analog signals are transmitted from the front-end electronics, at post-acceleration potential, to the grounded data-acquisition system via plastic optical fibers.

\subsubsection{Data acquisition}
\label{sec:daq}
The KATRIN data-acquisition (DAQ\nomenclature{DAQ}{Data acquisition}) systems used to collect FPD (\sref{sec:fpd}) and monitor-spectrometer (\sref{sec:mos}) data utilize custom first-level trigger (FLT\nomenclature{FLT}{First-level trigger}) and second-level trigger (SLT\nomenclature{SLT}{Second-level trigger}) hardware. As described in detail elsewhere~\cite{Amsbaugh2015}, each FLT card is equipped with 24 channels of analog signal conditioning and processing, including analog differential receivers with programmable offsets; programmable amplifiers; bandpass filters; digitizer drivers; serial analog-to-digital converters (ADCs\nomenclature{ADC}{Analog-to-digital converter}) with 12-bit precision and a 20-MHz sampling rate; and auxiliary memory for the ADCs. Each system has an SLT card used to read out data from the FLT cards, and communicates with a primary DAQ computer via a Gigabit Ethernet interface. The SLT cards provide timing signals that synchronize the internal counters of the DAQ electronics, allowing searches for coincident events even between different systems. On-board field-programmable gate arrays (FPGAs\nomenclature{FPGA}{Field-programmable gate array}) control acquisition and preprocessing while an additional central-control FPGA performs synchronization and readout for each card.

The FLT FPGAs support three different data-collection modes. The primary data-taking mode, energy mode, records energy and timing for each event, as reconstructed online with a cascade of two trapezoidal filters~\cite{jor94,Amsbaugh2015}. Trace mode additionally records the 2048-point ADC waveform of each event. In histogram mode, designed for very high rates, the FLT hardware fills a 2048-bin energy histogram. By using the FLT's 64-page, 2048-word ring-buffers, deadtime-free energy-mode rates of up to \SI{108}{\kilo cps} can be acquired over the whole detector.

The DAQ hardware is controlled and read out using Object-oriented Real-time Control and Acquisition (ORCA\nomenclature{ORCA}{Object-oriented Real-time Control and Acquisition data-acquisition software}) software~\cite{Howe2004}. ORCA, written in Objective-C for the Mac OS X operating system, uses object-oriented programming techniques to encapsulate hardware elements and DAQ concepts into software objects. A graphical user interface allows the operator to configure and run the KATRIN hardware. ORCA is fully scriptable and custom scripts were used extensively during data collection.

\subsection{Ion and electron sources}
\label{sec:sources}
The four commissioning sources used in this work, summarized in \tref{SourceSummary}, are located in several positions along the beamline (\fref{fig:beamline}). Moving from the rear toward the focal-plane detector, the ELIOTT (ELectron impact IOn source To Test the DPS\nomenclature{ELIOTT}{ELectron impact IOn source To Test the DPS}) ion source (\sref{sec:eliott}) and photoelectron source (\sref{sec:rw_pe}), both used in the First Light campaign (\sref{sec:FirstLight-campaign}) were part of the rear section (\sref{sec:rearsection}). The Gaseous \kr{} Source (\sref{sec:GKrS}), used for the measurement campaign described in \sref{sec:GKrS-campaign}, was introduced into the windowless, gaseous source section (\sref{sec:wgts}). The Condensed \kr{} Source (\sref{sec:CKrS}), used for the measurement campaign described in \sref{sec:CKrS-campaign}, was deployed in the cryogenic pumping section (\sref{apparatus:transport}).

\begin{table}[tbp]
	\caption{\label{SourceSummary}Beamline commissioning sources used in this work}
	\begin{center}
		\begin{tabular}{llcll}
			\hline
			\textbf{Source} & \textbf{Location} & \textbf{Particle} & \textbf{Energy (eV)} &  \textbf{Measurement} \\
			\hline
			ELIOTT 	& Rear sec. &  D$_n^+$ & 2, 30, or 90 & First Light (\sref{sec:FirstLight-campaign}) \\
							& 					    & $\mathrm{e}^-$ & 100 & First Light (\sref{sec:FirstLight-campaign}) \\
			\hline
			Rear-wall photo\-electrons & Rear sec.  & $\mathrm{e}^-$ & $100-110$ & First Light (\sref{sec:FirstLight-campaign}) \\
			\hline
			Gaseous \kr{} & WGTS & $\mathrm{e}^-$ & $7000-32000$ & GKrS (\sref{sec:GKrS-campaign}) \\
			\hline
			Condensed \kr{} & CPS & $\mathrm{e}^-$ & $7000-32000$ & CKrS (\sref{sec:CKrS-campaign}) \\
			\hline
		\end{tabular}
	\end{center}
\end{table}

\subsubsection{ELIOTT ion source}
\label{sec:eliott}
\begin{figure}[tbp]
	\centering
	\begin{minipage}{0.55\textwidth}
		\centering
		\includegraphics[width=\textwidth]{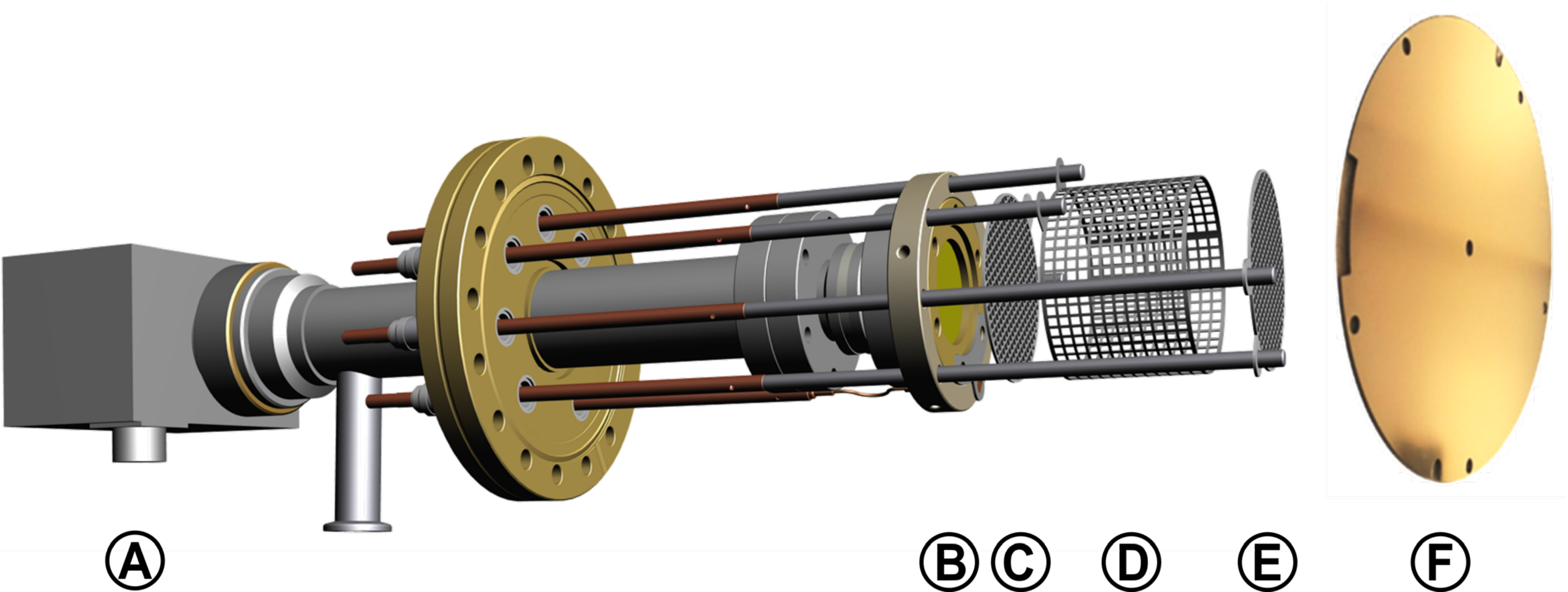}
	\end{minipage}
	\begin{minipage}{0.4\textwidth}
		\centering
		\includegraphics[width=.9\textwidth]{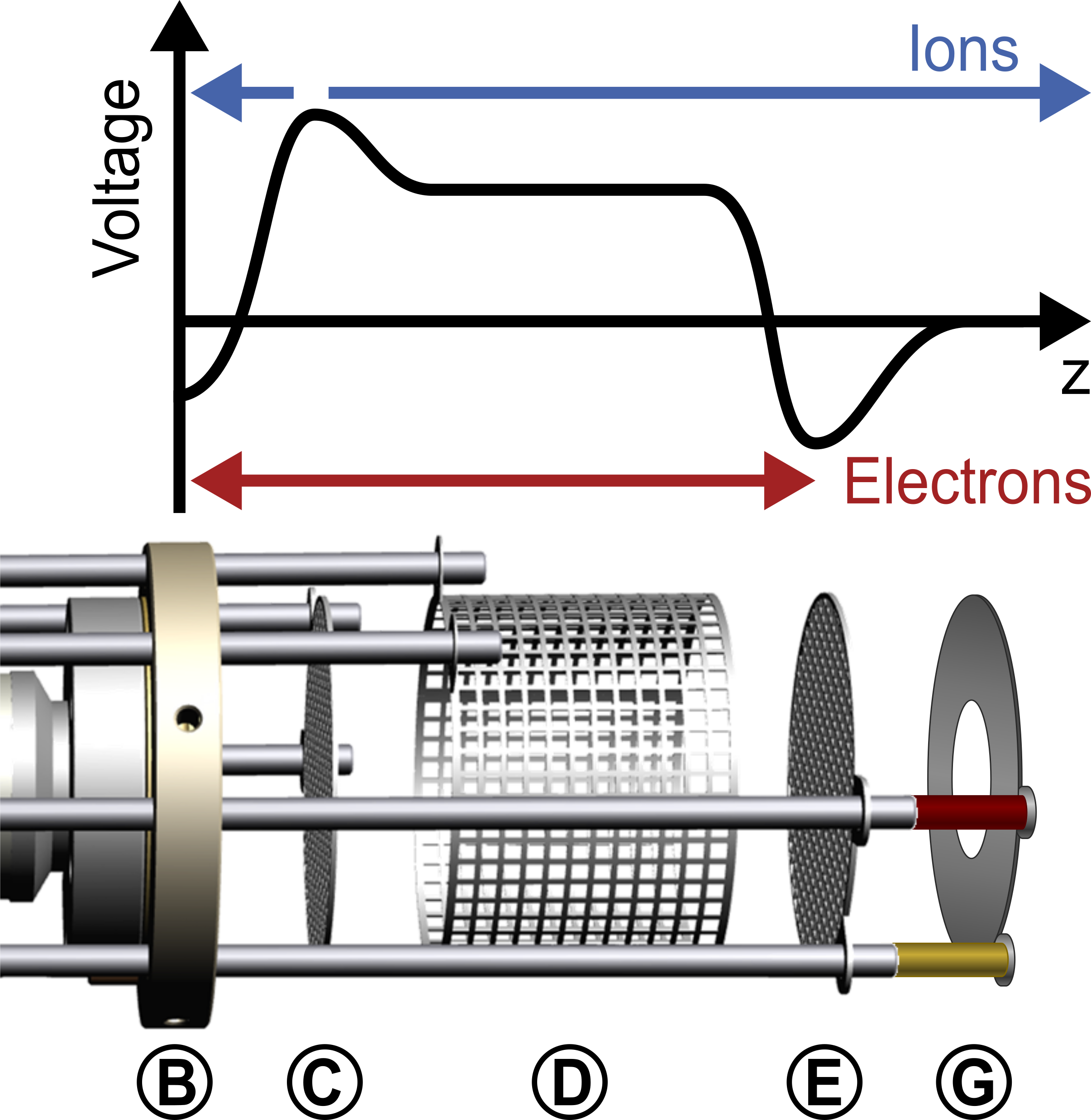}
	\end{minipage} 
	\caption{\label{pic.ELIOTT}\textbf{Drawing of the ELIOTT ion source.}
	Left: Engineering model of ELIOTT. The UV light from the lamp (A) is transmitted onto the photocathode window (B), which is gold-coated on the vacuum side and emits electrons. Right: Schematic plot of voltage variation along the ELIOTT apparatus. The photoelectrons are removed from the window (B) at \SI{-1}{\volt} and stored with the electrode (E) at \SI{-7}{\volt}. Most of the ionization occurs inside electrode (D) at \SI{90}{\volt}. The ions are prevented from traveling upstream by the \SI{120}{\volt} at electrode (C), and are partly detected with the ring-shaped Faraday cup (G) at ground potential. The model shows a shorter distance between ELIOTT and the rear wall (F) than the actual \SI{1}{\metre} mounting distance during the measurement campaign.}
\end{figure}

The ELIOTT ion source~\cite{Lukic2011} was mounted inside the bore of the rearmost superconducting magnet, behind the rear wall (\sref{sec:rearsection}). 
D$_2$ gas was provided as a small, almost constant flow from a buffer vessel. The gas was ionized with photoelectrons by means of the electrode setup depicted in \fref{pic.ELIOTT}. A \SI{5}{\milli\metre} aperture collimated the ions into a narrow pencil beam entering the beamline. The ions were presumably D$^+$, D$_2^+$ and D$_3^+$, but a quantitative understanding of the ion-beam makeup will require further work modeling the pressure-dependent ion-transformation processes. 

The deuterium pressure could not be measured inside ELIOTT due to space constraints, and must be extrapolated from measurements in the pressure supply. For the ionization procedure described in \fref{pic.ELIOTT}, the D$_2$ pressure was estimated at about \SI{1.2e-3}{\milli\bar}, \SI{2e-2}{\milli\bar} and \SI{4e-2}{\milli\bar}, creating deuterium ions with energies peaking at about \SI{90}{\electronvolt}, \SI{30}{\electronvolt} and \SI{2}{\electronvolt}, respectively. The \SI{2}{\electronvolt} thermal ions were generated with the ELIOTT hull and the rear wall set to a slightly positive potential to provide an energy offset.

An electron pencil beam could be produced instead of an ion beam by evacuating ELIOTT and grounding all electrodes. The photocathode window was set to be more negative than \SI{-100}{\volt} to supply the electrons with enough energy for transport. 

\subsubsection{Rear-wall photoelectron source}
\label{sec:rw_pe}
\begin{figure}[tbp]
	\centering
	\includegraphics[width=0.8\textwidth]{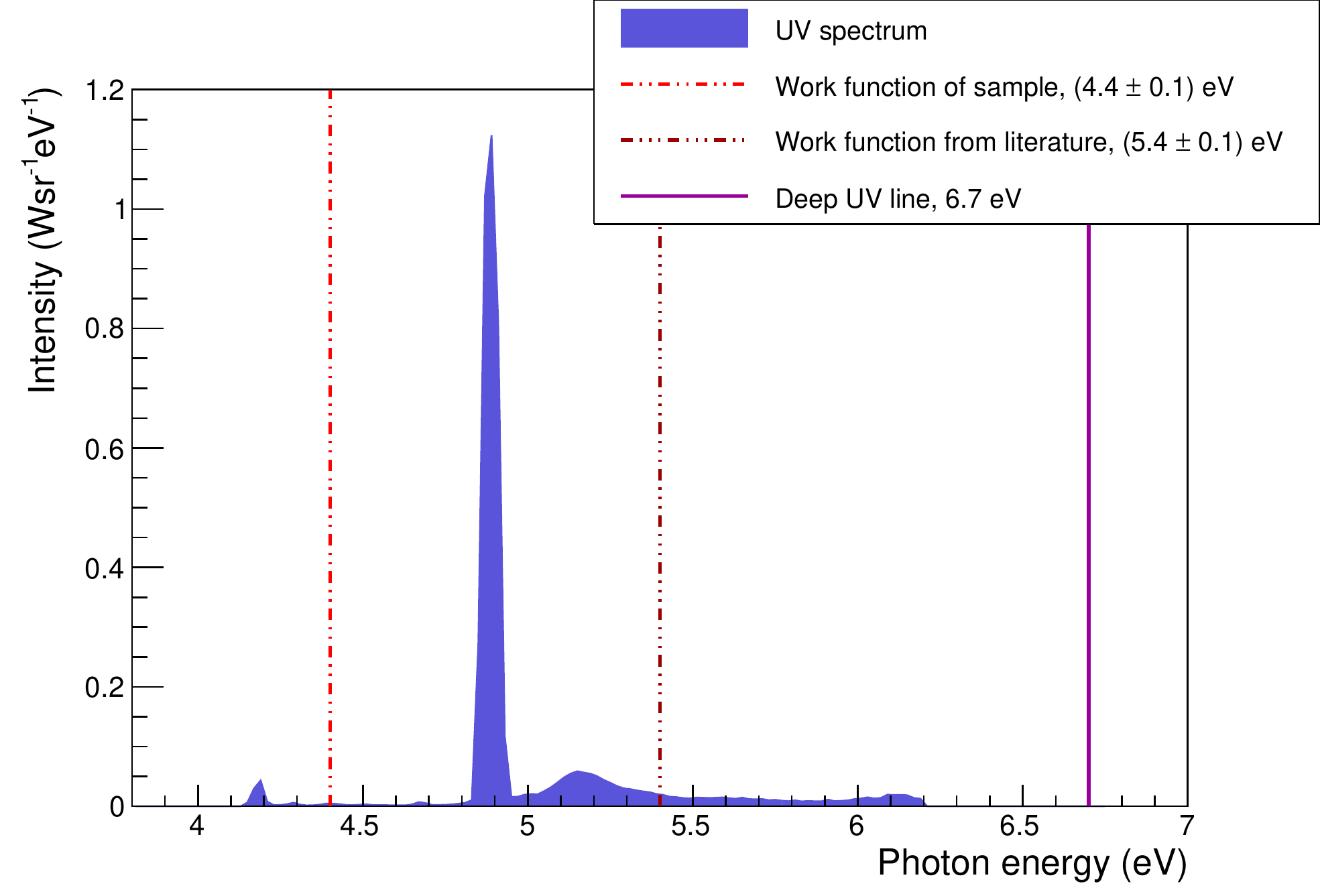}
	\caption{UV part of the spectrum (\num{200}$-$\SI{300}{\nano\metre}) of the rear-wall illumination source. Most of the intensity is well above \SI{4.4}{\electronvolt}, the measured rear-wall work function. The solid line indicates the location of an expected deep-UV line, outside the measurement range but confirmed via commissioning measurements with the monitor spectrometer (\sref{sec:mos}).}
	\label{fig:Rearwall-UVspectrum}
\end{figure}

For the First Light measurements (\sref{sec:FirstLight-campaign}), electrons were generated by illumination of the rear wall with UV light. A mini Z\footnote{RBD Instruments, Bend, Oregon 97701 USA.} water desorption system was used as a photon generator with a peak intensity at about \SI{253}{\nano\metre} (\SI{4.9}{\electronvolt}), as shown in the spectral measurement in \fref{fig:Rearwall-UVspectrum}. This is well above the work function of \num{4.3}--\SI{4.5}{\electronvolt} of our samples, but is lower than values of \num{5.3}--\SI{5.5}{\electronvolt} reported for single crystal surfaces~\cite{Lide2000}. For KATRIN it is most important that the work function is stable in time and homogeneous over the full area of \SI{140}{\centi\metre\squared}~\cite{PhDSchoenung2016}. 

Electrons emitted by the photo\-electric effect were accelerated by a bias voltage up to \SI{500}{\volt} for transport through the beamline; the electrons were raised above the FPD detection threshold by the \SI{10}{\kilo\volt} post-acceleration potential (\sref{sec:fpd}). To reduce the electron rate, the mini Z light was attenuated via a long, evacuated light path with two \SI{90}{\degree} elbows, which guided the light into the rear chamber through a sapphire window. 

\subsubsection{Gaseous \kr{} source}
\label{sec:GKrS}
The Gaseous \kr{} Source (GKrS\nomenclature{GKrS}{Gaseous \kr{} source}) is based on deposition of the parent radionuclide \textsuperscript{83}Rb into zeolite (molecular sieve) beads. Long-term commissioning with the short-lived \kr{} (T$_{1/2}$ = \SI{1.8}{\hour}) is simplified since the krypton can be continuously supplied via the decay of the parent \textsuperscript{83}Rb, which has a favorable half-life of \SI{86.2}{days}. For KATRIN the \textsuperscript{83}Rb is produced at the NPI \v{R}e\v{z} cyclotron in the reactions \textsuperscript{nat}Kr(p,xn)\textsuperscript{83}Rb, using \SI{27}{\mega\electronvolt} protons at \SI{15}{\micro\ampere} beam current. After irradiation, the \textsuperscript{83}Rb, settled on the target walls, is washed out by de-ionized water and deposited by sorption into 15 zeolite beads (Merck, type 5A)~\cite{Venos2005}. From such a zeolite source, \SIrange{80}{90}{\percent} of the produced \kr{} emanates into the vacuum at room temperature~\cite{Venos2014}. Meanwhile, the \textsuperscript{83}Rb is firmly bound in the zeolite beads, minimizing contamination with long-lived isotopes. The \textsuperscript{83}Rb activity in two sets, each of 15 zeolite beads, amounted to \SI{1.0}{\giga\becquerel} on the first day of the gaseous \kr{} campaign (\sref{sec:GKrS-campaign}).

\begin{figure}[tbp]
	\centering
	\includegraphics[width=11cm]{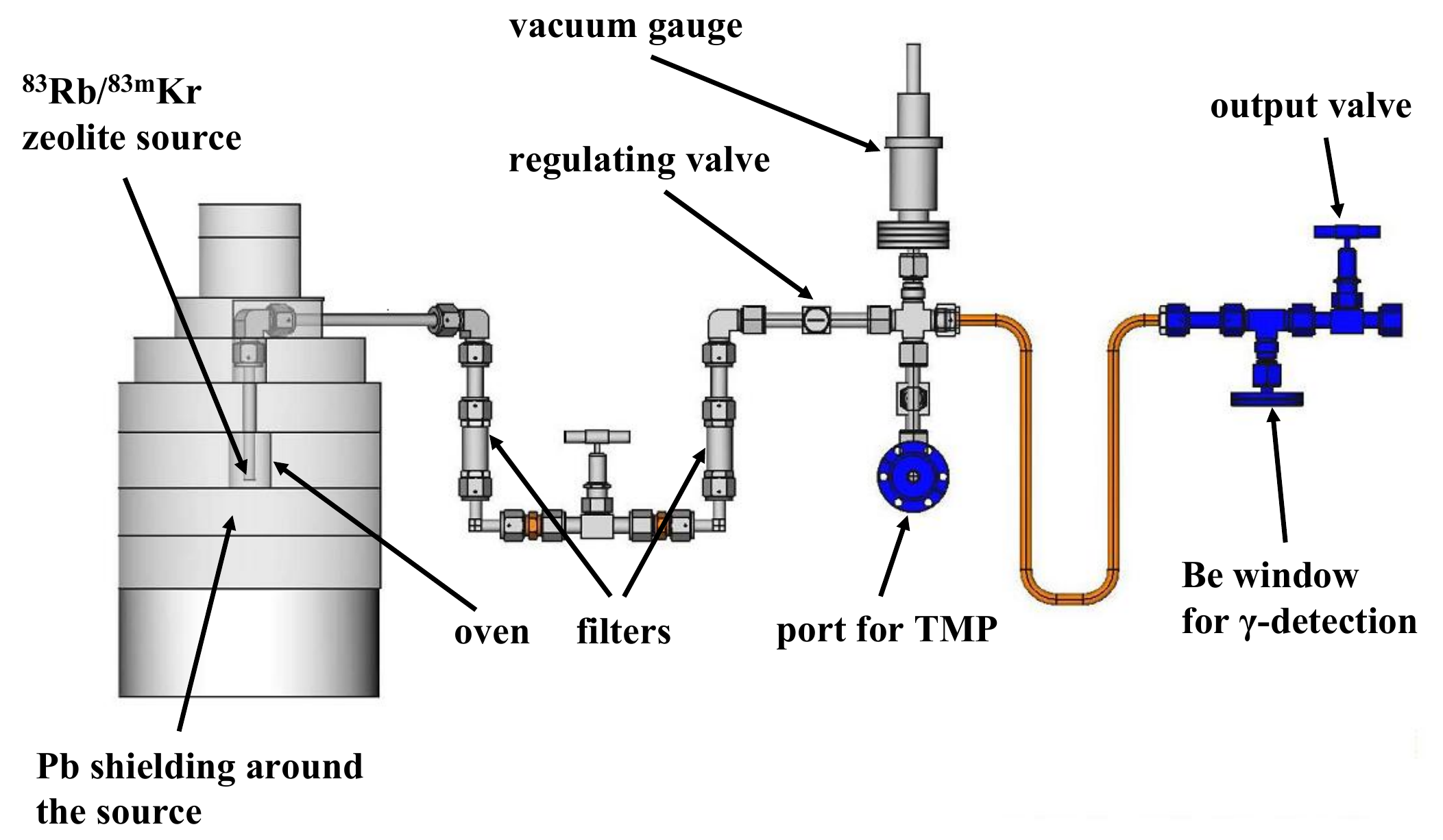}
	\caption{Schematic diagram of the \textsuperscript{83m}Kr generator.}
	\label{fig:generator}
\end{figure}

The gaseous krypton generator~\cite{Sentkerestiova2017,Sentkerestiova2017b}, shown in \fref{fig:generator}, injects the emanated \kr{} into the WGTS through a pumping port. The generator setup uses stainless-steel VCR components. The \textsuperscript{83}Rb-loaded zeolite beads sit in the bottom of a well embedded in a cylindrical oven, which is in turn situated in demountable lead shielding. In this way, the source beads can be baked in order to desorb residual gases that would otherwise affect the WGTS vacuum. Two sintered filters with \SI{0.5}{\micro\metre} pores prevent aerosols and small zeolite abrasions, which might contain \textsuperscript{83}Rb, from diffusing into the WGTS. A Be window allows detection of $\upgamma$ and X-rays from the \kr{} decay by means of a silicon drift detector. A regulating valve controls the amount of \kr{} gas injected into the WGTS. Just before use and during baking, the generator volume is evacuated through the bellows with a TMP. 

\subsubsection{Condensed \kr{} source}
\label{sec:CKrS}
\begin{figure}[tbp]
	\centering
	\includegraphics[width=.5\textwidth]{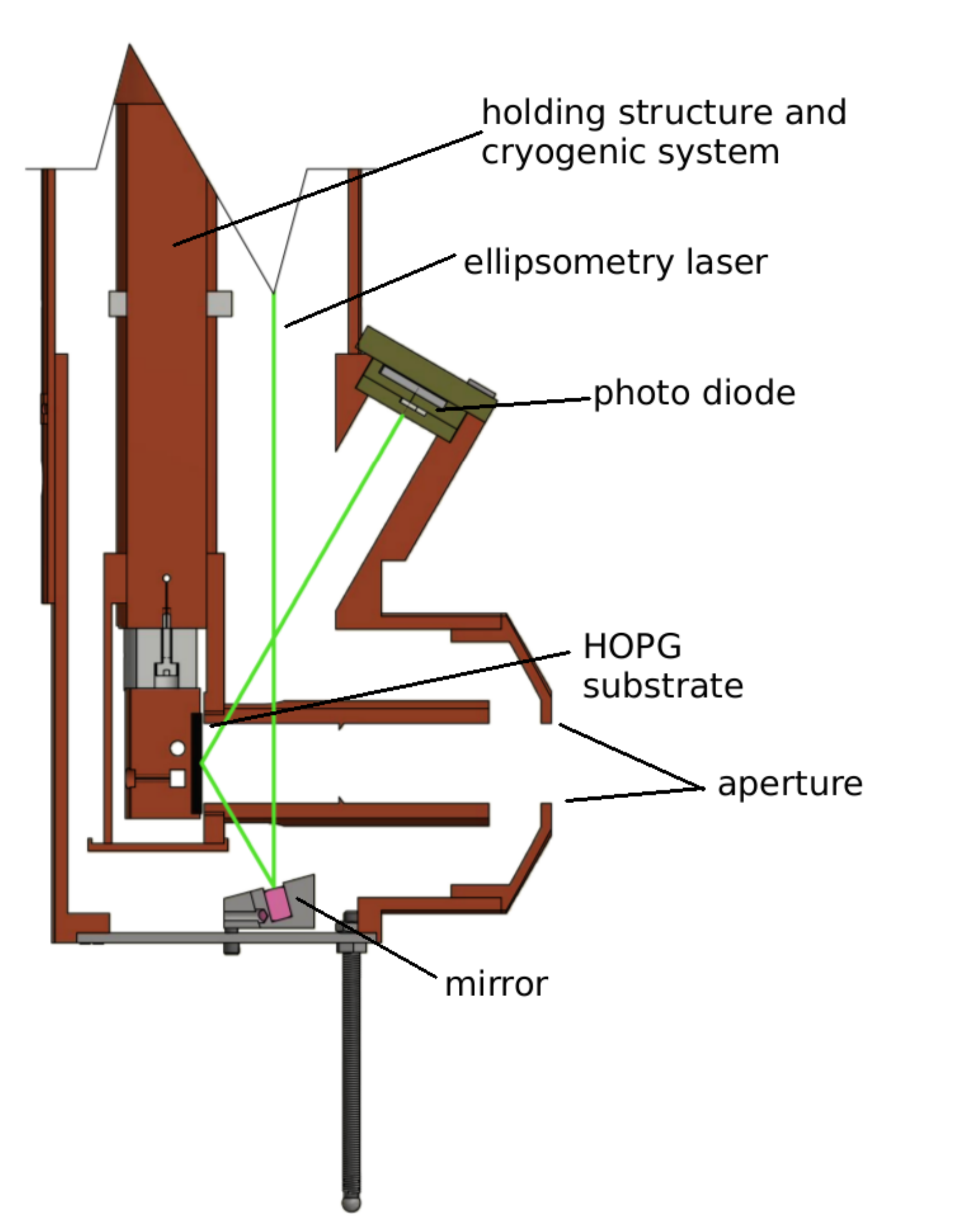}
	\caption{Schematic view of the lower part of the CKrS apparatus around the substrate, including the path of the ellipsometry laser beam. The setup can be moved up and down, thus entering or retracting from the beamline. Additionally it can be tilted perpendicular to the illustrated plane, enabling the CKrS to scan the whole flux tube.}
	\label{fig:ckrs_geometry}
\end{figure}

The Condensed \kr{} Source (CKrS\nomenclature{CKrS}{Condensed \kr{} Source}) is a thin sub-monolayer film of radioactive \textsuperscript{83m}Kr condensed on a highly oriented pyrolytic graphite (HOPG\nomenclature{HOPG}{Highly oriented pyrolytic graphite}) substrate with a diameter of \SI{2}{\centi\metre}, cooled to \SI{26}{\kelvin}. The gaseous \kr{} is generated from decaying \textsuperscript{83}Rb atoms bound in zeolite beads, and then transferred by diffusion through a gas system and a heatable capillary towards the substrate. The HOPG substrate is shielded from condensing residual gas by an inner cold shield at about \SI{16}{\kelvin} and an outer cold shield at less than \SI{100}{\kelvin}. An overview of the CKrS geometry is shown in \fref{fig:ckrs_geometry}.

In addition to the source and condensation system, the CKrS includes a laser ellipsometry system~\cite{Bauer2013b} to measure the thickness of the layer frozen onto the substrate. This might either be residual gas or stable krypton, which can be fed into the gas system for ellipsometry tests and pre-plating of the substrate. Before condensing a film of radioactive krypton onto the substrate, it is cleaned of residual-gas atoms by heating it to \SI{120}{\kelvin} via diodes and applying a high-powered frequency-doubled Nd:YAG ablation laser beam.

The CKrS is located in the CPS (\sref{apparatus:transport}), between the source and the detector, so it must be moved outside the flux tube during normal KATRIN operation. The entire gas-handling and ellipsometry apparatus is therefore mounted on a carriage, which can be moved in the vertical direction along sliding rails. Horizontal motion is achieved by tilting the sliding rails, with their hinges serving as pivot point.

For future measurements, the CKrS can be placed at a potential of up to \SI{-1}{\kilo\volt} in order to shift the electron energy of the K line from \SI{17.8}{\kilo\electronvolt} to the \SI{18.6}{\kilo\electronvolt} endpoint energy of the tritium \betaspec{}. Further details of the CKrS and its subsystems can be found in Ref.~\cite{PhDBauer2014}.

\subsection{Calibration and monitoring}
\label{sec:calmonitor}
Various calibration and monitoring components are deployed in the KATRIN experiment. Here we discuss the high-voltage subsystem (\sref{par:high-voltage}) with its parallel monitor-spectrometer calibration setup (\sref{sec:mos}), an X-ray detector for monitoring source activity (\sref{apparatus:BIXS}), and a detector for monitoring particles traveling in the flux tube (\sref{apparatus:FBM}). The slow-control system (\sref{sec:slowcontrols}) provides control, monitoring and data management over the entire beamline.

\subsubsection{High-voltage subsystem}
\label{par:high-voltage}
\begin{figure}[tbp]
	\centering
	\includegraphics[width=0.75\textwidth]{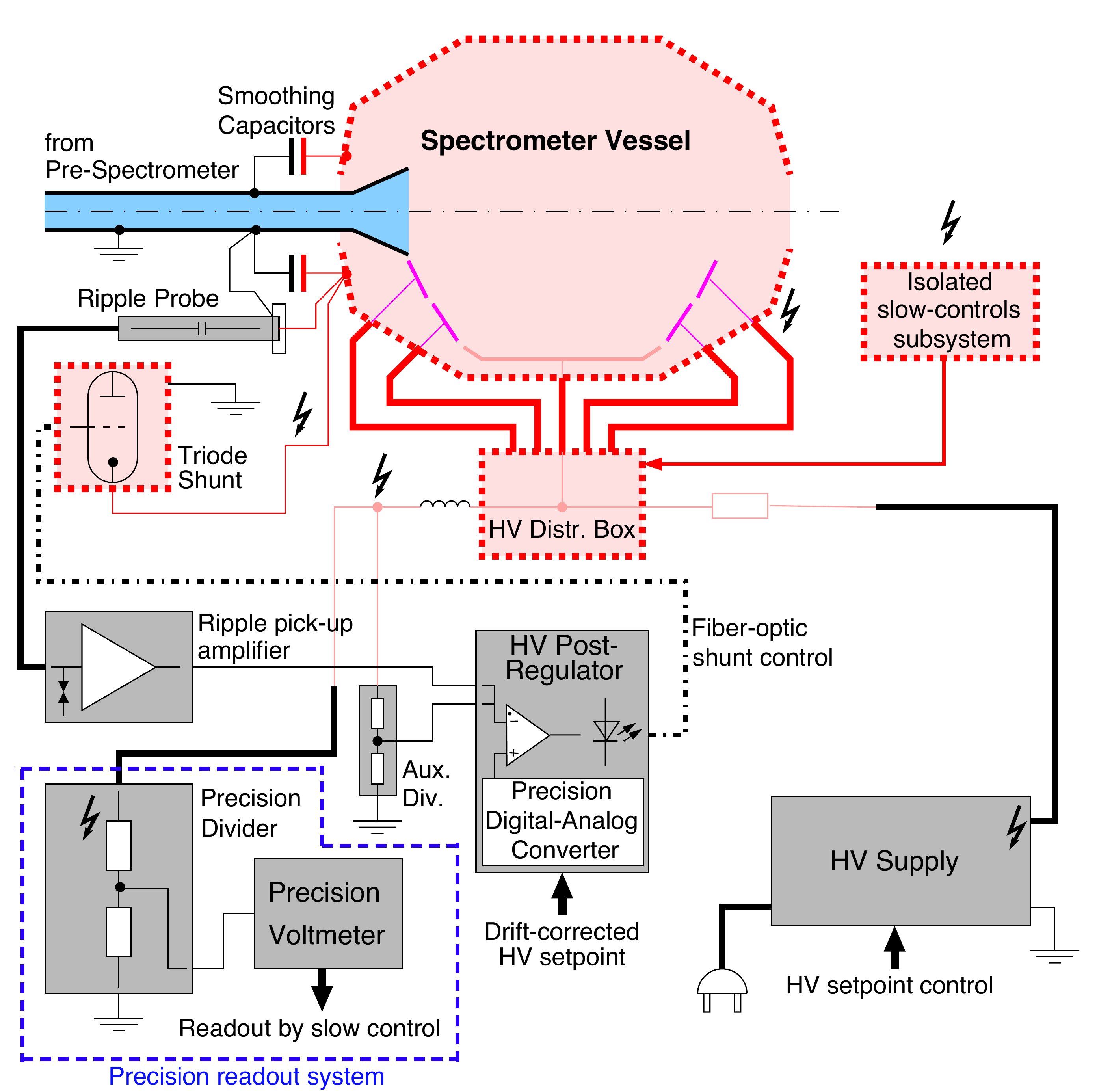}
	\caption{Block diagram of the high-voltage subsystem, including precision DC monitoring as well as AC ripple readout and post-regulation. Thick dotted lines surround components at the retarding potential.
	\label{fig:high_voltage}}
\end{figure}

The retarding high-voltage supply of the KATRIN main spectrometer (\sref{apparatus:sds}) is crucial to the overall accuracy of the experiment. As defined in the KATRIN design report~\cite{KATRIN2005}, allowed deviations of the high voltage are at the \SI{3}{ppm} level, i.e.\ \SI{60}{\milli\volt} at \SI{-18.6}{\kilo\volt}. This accuracy requirement applies even at very high frequencies, due to the short electron transit time through the spectrometer.

The high-voltage system must provide the spectrometer vessel with a constant retarding voltage, while maintaining the whole KATRIN beamline as a Faraday cage with respect to any alternating-current (AC\nomenclature{AC}{Alternating current}) ripple. \figref{fig:high_voltage} shows a block diagram of the system. Precision voltage stabilization begins with a primary, commercial high-voltage supply unit\footnote{HCP 70M-35000, custom-made by FuG with improved stability and noise, higher programming resolution, and an interlock system.} which is stabilized to a \num{e-4} ($\sim$~\SI{1}{\volt}) accuracy and feeds the spectrometer vessel via a series resistor of about \SI{22}{\kilo\ohm}. Fast and precise fine regulation is done by a vacuum triode shunt regulator (ballast) connected directly across the spectrometer entrance insulator, minimizing stray inductance. The triode shunt is controlled by a precision regulator, which obtains its setpoint from a \SI{20}{bit} digital-to-analog converter. The acquisition of the actual high-voltage value is divided into two signal paths: (1) A fast, non-attenuating AC path consisting of a high-voltage decoupling capacitor (the ripple probe), a protection network and a buffer amplifier, and (2) a slow DC path consisting of a precision voltage divider~\cite{Thuemmler2009, Bauer2013a}. The ripple probe is also connected directly across the entrance insulator and has an upper cutoff frequency on the order of \SI{1}{\mega\hertz}. The post-regulation system not only suppresses \SI{50}{\hertz} power-grid interference, but also radiofrequency interference from various sources. Frequencies higher than \SI{1}{\mega\hertz} are effectively shorted by three, \SI{120}{\degree}-spaced high-voltage filtering capacitors (\SI{7}{\nano\farad} each) across the spectrometer entrance insulator.

During the krypton measurement phases described in \sref{sec:GKrS-campaign} and \sref{sec:CKrS-campaign}, the filament heating transformer of the shunt triode exhibited excessive partial discharges. Therefore, the triode shunt had to be disconnected from the spectrometer at retarding voltages beyond \SI{-20}{\kilo\volt}. With the post-regulation loop disabled, the dominant component of AC interference on the high voltage was a \SI{50}{\hertz} sine wave of roughly \SI{400}{\milli\volt} amplitude (peak-to-peak) caused by pickup from the power grid, which must be included into the data analysis model. The post-regulation system will be retrofitted with an improved transformer.

\subsubsection{Monitor spectrometer}
\label{sec:mos}
As a complementary technique for monitoring the retarding voltage, in parallel to the KATRIN main beamline, the MAC-E-type monitor spectrometer~\cite{Erhard2014} observes mono-energetic conversion electrons emitted from a nuclear standard, the implanted \kr{} source. The relative stability of the measured line energy has been shown to be better than ppm-level over the course of several weeks~\cite{Zboril2013, PhDSlezak2015}. This allows continuous monitoring of the high-voltage stability throughout KATRIN operations. A significant change of the line energy detected by the monitor spectrometer might indicate an instability of the high-voltage measurement.

The monitor spectrometer is a stainless-steel vessel, \SI{3}{\meter} long and \SI{0.5}{\meter} in radius, equipped with stainless-steel retarding electrodes and super- and normal-conducting coils to create and shape the guiding magnetic field. The energy resolution of \SI{0.93}{\electronvolt} for the \SI{18.6}{\kilo\electronvolt} electrons is similar to the energy resolution of the main spectrometer. The retarding electrode systems of the monitor and main spectrometer are connected, allowing the same high voltage to be applied at both spectrometers. To bridge the gap of about \SI{750}{\electronvolt} between the K-32 conversion-electron line energy and the tritium \betadec{} endpoint, the implanted \kr{} source is biased by a dedicated power supply which is read out using a commercial voltmeter.

\begin{figure}[tbp]
	\centering
	\includegraphics[width=12cm]{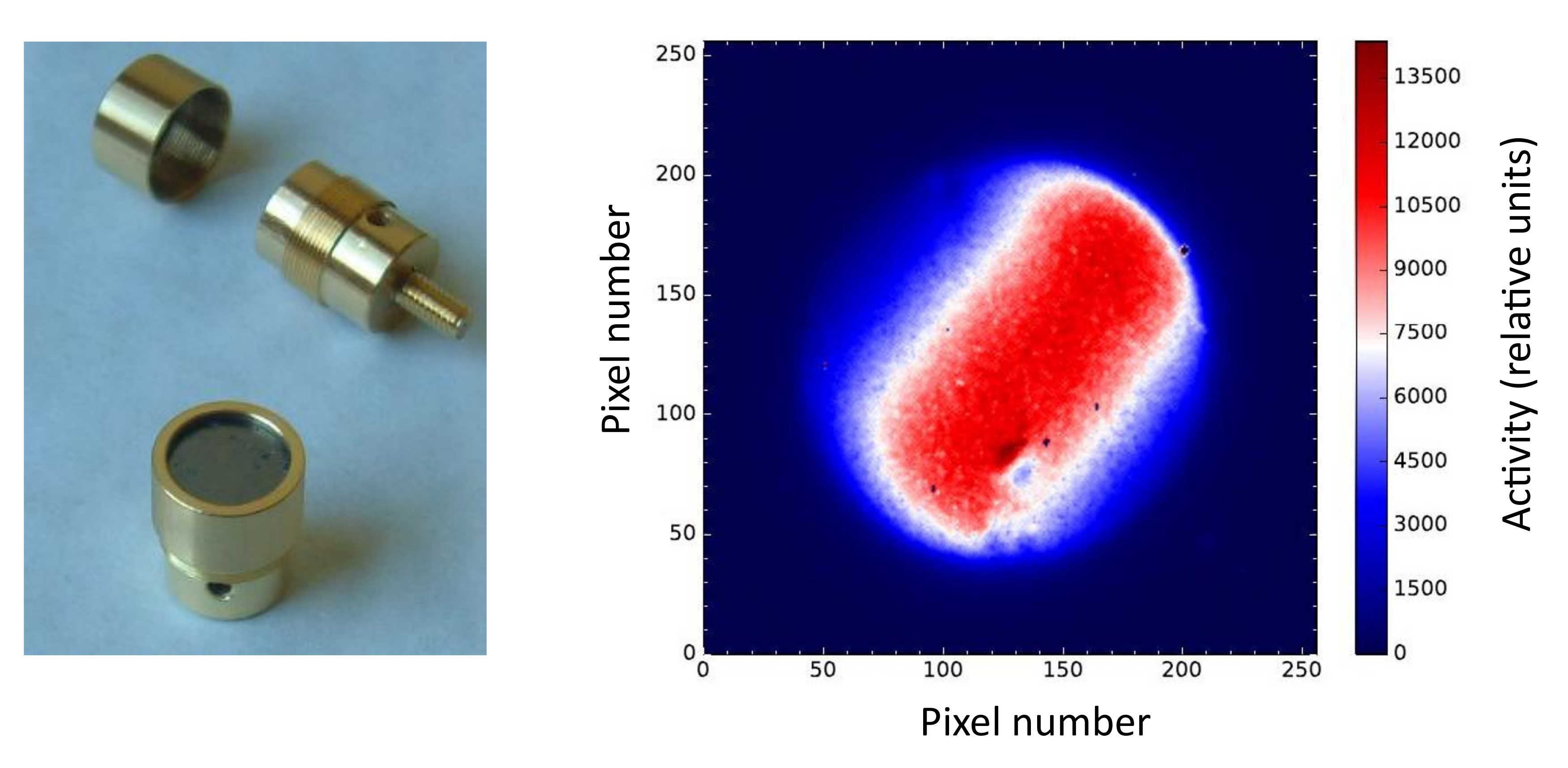}
	\caption{Left: The implanted \kr{}/\textsuperscript{83}Rb source mounted in the gold-coated holder. Right: A two-dimensional source activity map measured with a \num{14x14}~\SI{}{\milli\metre\squared} Timepix detector.}
	\label{fig:imkrs}
\end{figure}

The electron source for the monitor spectrometer is produced by implanting \textsuperscript{83}Rb ions into a HOPG foil~\cite{PhDZboril2011,Zboril2013,Erhard2014,PhDSlezak2015,arenz:phd2017}. Repeated measurements~\cite{Zboril2013} revealed that the spectral line positions of the electrons emitted from the source change slowly with time. This very small effect is probably connected with the ageing of the source and with changes of the spectrometer workfunction. This energy drift is less than \SI{20}{\milli\electronvolt} per month for a concentration less than \num{6e20} $^{83}$Rb ions \SI{}{\per\centi\metre\cubed} in the source~\cite{PhDSlezak2015}. A very similar linear dependence of energy on the time was also observed for the strong L$_{3}$-32 line,  which in turn can be used for high-voltage monitoring near \SI{30}{\kilo\volt}.

The source (\fref{fig:imkrs}) was prepared at the Bonn University mass separator using \textsuperscript{83}Rb produced at the NPI \v{R}e\v{z} cyclotron, using a HOPG foil \SI{12}{\milli\metre} in diameter and \SI{0.5}{\milli\metre} thick. The foil normal was tilted by \SI{10}{\degree} relative to the \SI{8}{\kilo\electronvolt} mass-separator ion beam in order to reduce the contribution of channeling effects to the implantation depth. The \textsuperscript{83}Rb activity was \SI{4.8}{\mega\becquerel}, as measured with a silicon drift detector. The two-dimensional distribution of \textsuperscript{83}Rb  was measured using a \num{14x14}~\SI{}{\milli\metre\squared} Timepix detector~\cite{Jakubek2009}, as shown in \fref{fig:imkrs}. Combined with the depth distribution of the \textsuperscript{83}Rb ions calculated with SRIM~\cite{Ziegler2010}, we computed an average volume concentration of \SI{3.8e20}{ions\per\centi\metre\cubed}. 

\subsubsection{Beta-induced X-ray spectroscopy system}
\label{apparatus:BIXS}

KATRIN will utilize a beta-induced X-ray spectroscopy (BIXS\nomenclature{BIXS}{Beta-induced X-ray spectroscopy}) system to measure the activity of the tritium gas column inside the WGTS (\sref{sec:wgts}), with sensitivity to \SI{0.1}{\percent} fluctuations within approximately \SI{100}{\second}~\cite{PhDRoellig2015,Roellig2015,Babutzka2012}. When electrons from tritium decay in the source section hit the rear wall (\sref{sec:rearsection}), they produce X-rays via scattering interactions. These X-rays are observed in the BIXS systems. For the GKrS measurements (\sref{sec:GKrS-campaign}) the setup was adapted and simplified. A single BIXS detector was deployed on-axis with the WGTS beam tube at its rear side. A gold-coated Be window of \SI{250}{\micro\metre} total thickness with a \SI{100}{\nano\metre} Au layer and a \SI{<10}{\nano\metre} Ti intermediate layer was placed in front of the \SI{100}{\milli\metre\squared} windowless silicon drift detector\footnote{KETEK AXAS M with Amptek DP5 DPP.} to prevent it from contamination. With this setup the source activity can be monitored on time scales of order \SI{100}{\second}. Characteristic X-ray lines in the spectrum enable calibration of the system in situ. 

\subsubsection{Forward beam monitor}
\label{apparatus:FBM}
The Forward Beam Monitor (FBM\nomenclature{FBM}{Forward Beam Monitor})~\cite{ell17,Babutzka2012,PhDSchmitt2008} is used in KATRIN for monitoring the relative intensity of the electron flux just before it enters the spectrometers (\sref{apparatus:sds}), as shown in \fref{fig:beamline}. The FBM can measure the \betael{} flux with a precision of \SI{0.1}{\percent} in less than \SI{60}{\second}. $\upbeta$-spectra are measured for later analysis.

\begin{figure}[tbp]
	\centering
	\includegraphics[width=1\textwidth]{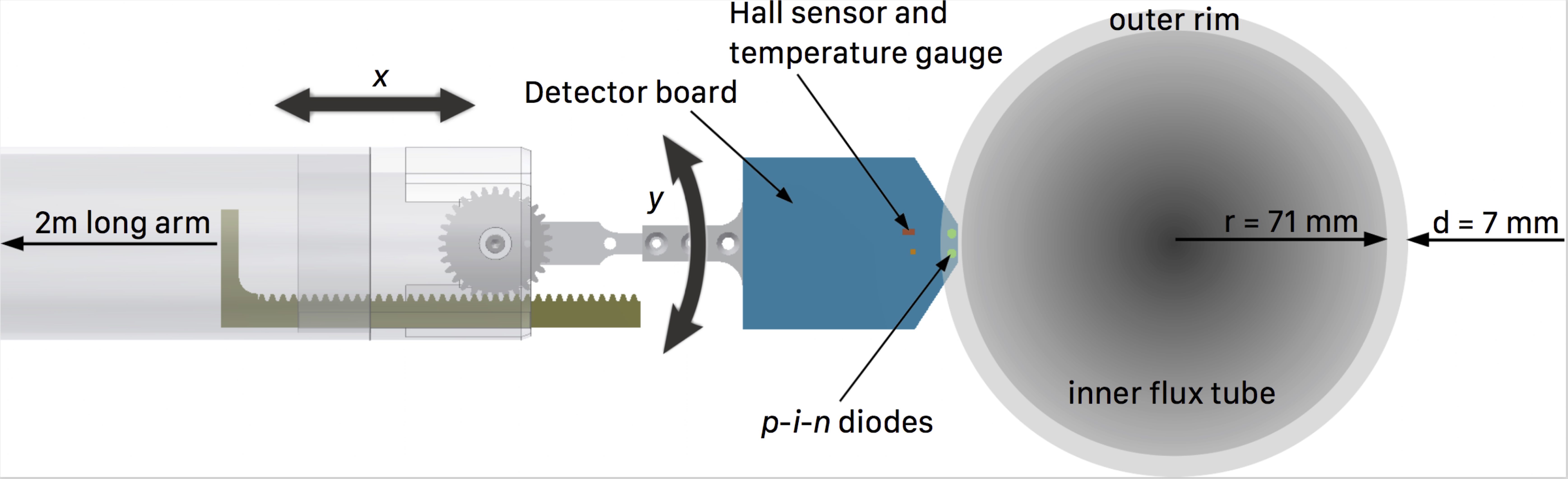}
	\caption{Two independent motion systems drive the FBM detector board around the entire cross section of the flux tube in the CPS. The first system moves the front end linearly along the $x$-direction and the second performs rotational movement. When the FBM is in standard monitoring position, it sits at the outer rim of the flux tube and continuously measures the flux properties without shadowing the FPD.}
	\label{fig:FBM_front-end}
\end{figure}

A vacuum manipulator with a \SI{2}{\metre} long bellows enables the FBM detector board to be inserted directly into the flux tube in the CPS, at any radial position inside the beam tube (as shown in \fref{fig:FBM_front-end}). The FBM can be moved with a precision of better than \SI{0.1}{\milli\metre} relative to its starting position. During nominal KATRIN runs the detector will sit at the outer rim of the flux tube, continuously monitoring the flux properties without shadowing the FPD. During the gaseous \kr{} measurement phase the FBM was located slightly further inside the flux tube (see \sref{subsec:GKrS:first-electrons}).

The detector board is equipped with two silicon \pin{} diodes; the diodes used during the gaseous \kr{} measurement phase had active areas of \SI{0.5}{\milli\metre\squared} and \SI{0.008}{\milli\metre\squared}. The energy resolution is approximately $\sigma_{\mathrm{FWHM}} = $ \SI{2}{\kilo\electronvolt} for \kr{} and tritium electron energies. The energy threshold during the gaseous \kr{} measurement phase was approximately \SI{16}{\kilo\electronvolt}, and for nominal tritium runs the energy threshold will be lowered to achieve high count rates for monitoring. The board also includes a temperature gauge and a Hall sensor for additional monitoring of the flux tube properties.

\subsubsection{Slow controls}
\label{sec:slowcontrols}
The KATRIN slow-control system is divided into three domains: controls, monitoring and data management. All three systems are continuously extended as new hardware components are added.

The control system ensures safe operation of critical components, such as the vacuum system, heating and cooling of the spectrometers, the cryogenic infrastructure and tritium management, and the source magnets. It uses programmable logic controller components from Siemens and is implemented in the PCS7 framework~\cite{pcs7-webpage} with a dedicated KATRIN library. It provides interlocks for other devices that belong to the monitoring domain, such as the detector or high-voltage systems. About 3100 sensors and actors are distributed over seven PCS7 stations and managed by three redundant PCS7 servers. Seventeen client stations are installed for operation. An Open PCS7 OPC server transfers about 2300 of these data streams for archival and usage in monitoring and data analysis. 

The monitoring domain integrates components essential for the precise control and characterization of the experimental status, such as the detector section, spectrometer magnets and magnetic-field monitoring, spectrometer temperature monitoring, high voltage, and various source calibration and monitoring systems. These normally have lower safety requirements and are enabled by interlocks from the control system where necessary; only these devices can be freely programmed by ORCA scripts (\sref{sec:daq}). The monitoring devices do not have strict automation-hardware requirements, but need to meet collaboration specifications. Most monitoring systems are based on standard automation components from National Instruments and are implemented with LabView~\cite{LabviewPage}. 

The slow-control data-management system continuously collects all relevant items for physics and technical analysis. With a delay of one or two minutes, all data are provided via web servers to the collaboration. The data management system uses the Advanced Data Extraction Infrastructure (ADEI\nomenclature{ADEI}{Advanced Data Extraction Infrastructure})~\cite{chilingaryan2010advanced, adei-webpage}. Intuitive graphical navigation enables both status checks and lookups of historical data. To improve performance, we use intelligent aggregation techniques to distill a few thousand data points from the millions that may span the time interval; caching techniques to speed data access; and asynchronous communication between server and user. The complete plot-generation time does not usually exceed 500 ms. The ADEI database is used to implement status displays for various components and measurement phases, and the system provides data-access patterns to correlate slow-control data and detector data. 

\section{First Light campaign}
\label{sec:FirstLight-campaign}

The First Light campaign was designed to demonstrate the electromagnetic operational readiness of the KATRIN apparatus. It was the first time that electrons were guided through the whole source and transport section into the spectrometers and onward to the detector. Here we used a simple setup in which the rear section (\sref{sec:rearsection}) was equipped with a gold-plated rear wall and simple UV illumination (\sref{sec:rw_pe}). A rear-wall potential up to \SI{-110}{\volt} accelerated the electrons into the WGTS (\sref{sec:wgts}), distributed radially and azimuthally in the beam tube. Instead of an electron gun, an ion source, ELIOTT, was used, providing pencil beams of electrons and deuterium ions (\sref{sec:eliott}). With three dipole electrodes in the DPS and four ring electrodes along the beamline (\sref{apparatus:transport}), the electric potential for ion drifting and blocking could be varied in a range up to \SI{380}{\volt}. Since the focus of this campaign was the transport of electrons and ions, it was sufficient to maintain the beamline vacuum with only a few TMPs along the transport section and a \SI{4.5}{\kelvin} CPS beam tube, resulting in an overall beamline pressure better than \SI{e-6}{\milli\bar}. The main spectrometer (\sref{apparatus:sds}), although unbaked, provided good vacuum conditions with a residual gas pressure on the order of \SI{e-9}{\milli\bar}. For most of the measurements, the source and pinch magnetic fields were set to $B_\mathrm{S} = \SI{0.72}{\tesla}$ and $B_\mathrm{PCH} = \SI{1.2}{\tesla}$, respectively, corresponding to \SI{20}{\percent} of their maximum values. Other magnetic fields along the beamline were scaled accordingly. The pre-spectrometer was maintained at ground potential.

\secref{sec:FirstLightFirstElectrons} describes the electrons seen in the detector during this campaign. \secref{sec:FirstLightAlignment} summarizes the alignment technique and results obtained, including the identification of an unexpected beamline blocking potential. \secref{sec:FirstLightBackground} gives the first KATRIN measurement of backgrounds induced by a large electron flux. \secref{sec:ion_detection} demonstrates ion transport and detection in the FPD. During neutrino-mass measurements, ions must be blocked or removed from the flux tube; \sref{sec:ionblocking} summarizes tests of active ion-blocking measures, while \sref{sec.unbeabsichtigteIonenblockierung} discusses the transport of thermal ions.

\subsection{First electrons}
\label{sec:FirstLightFirstElectrons}
The KATRIN experiment successfully reached a major milestone on 14 October 2016, when electrons were transmitted along the completed KATRIN beamline for the first time. The low-energy electrons were produced via the photo\-electric effect at the rear wall and were adiabatically guided to the FPD (\sref{sec:fpd}), where they were accelerated by a post-acceleration potential of \SI{10}{\kilo\volt} to rise above the detection threshold.

\begin{figure}[tbp]
	\centering
	\includegraphics[width=0.5\textwidth]{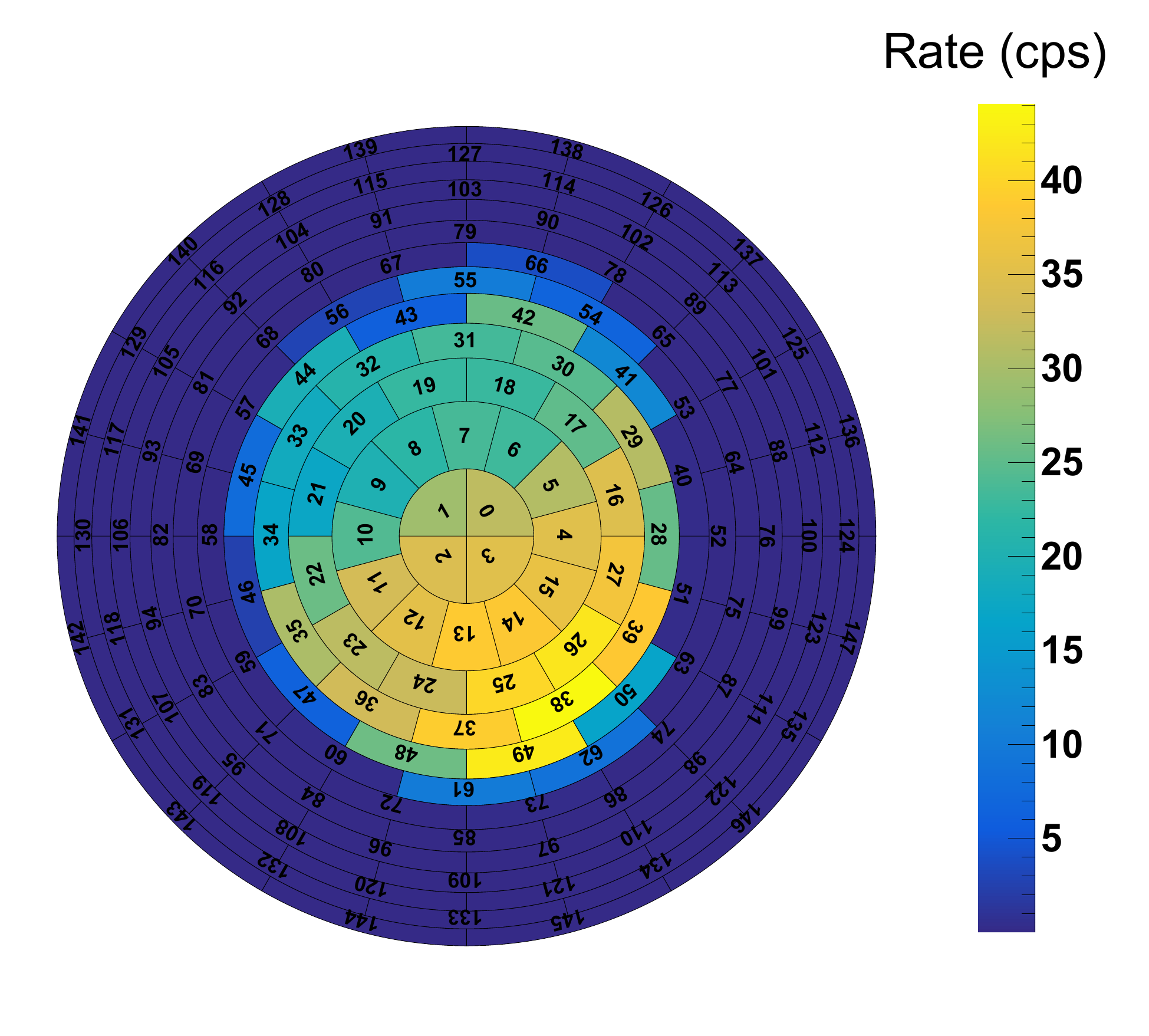}
	\caption{FPD pixel map of electrons originating from the rear wall. The color scale represents the rates of the individual detector pixels. In this unusual configuration, the illumination of the FPD was inhomogeneous and incomplete.}
	\label{pic.firstlight}
\end{figure}

For technical reasons one of the DPS magnets was not operational during this measurement, so the transported flux of electrons through the DPS was significantly reduced. The observed electron flux at the FPD was reduced by operating the detector magnet at \SI{1}{\percent} of its maximum value and therefore distributing the electron signal over several pixels (see \fref{pic.firstlight}). The electron beam was shifted inside the main spectrometer with the air coils~\cite{Erhard2017} in order to center the beam on the detector.

\subsection{Alignment and transport}
\label{sec:FirstLightAlignment}
Accurate alignment between the magnetic flux tube and the beamline minimizes electron loss between the source and the detector, and prevents the detection of electrons scattered on metal surfaces. The illuminated rear wall (\sref{sec:rw_pe}) served as an alignment electron source with a diameter of \SI{134}{\milli\metre}, covering the whole flux tube. Any collision between the flux tube and the beam tube would then cast a sharp shadow onto the detector, distinguishable from unknown inhomogeneities in the electron source. These shadows generally are localized in radius and in azimuth, but no axial information is available. To locate any collisions along the axis, the electron beam from ELIOTT (\sref{sec:eliott}) was used. The magnetic field was reduced section-wise, locally expanding the flux tube, and the electron beam was moved with the WGTS dipole coils until it collided with the beam tube in that section. The collision points were mapped onto the detector, and the radius of the flux tube passing through this section was calculated from the last illuminated pixel. Subpixel accuracy was achieved due to the linear dependence between the pencil-beam shift on the wafer and the dipole current. This method showed that there were no collisions of the nominal flux tube~\cite{PhDHackenjos2017}.
 
Not all electrons with energies below \SI{100}{\electronvolt} were transmitted from the rear wall to the detector. The blocking of these electrons was prominent in the lower part of the flux tube and showed a strong dependence on the electron energy. Below \SI{10}{\electronvolt} essentially no electrons were transmitted; above \SI{100}{\electronvolt} all electrons were transmitted. Between these energies, the rate of transmitted electrons increased with their energy.

A negative potential in the lower part of the beamline could explain this behavior. The axial position of this unintended blocking potential was narrowed down to the region between the pre-spectrometer and the main spectrometer. By applying a positive voltage to both ground electrodes and to the connecting beam-tube element, the shadowing of the low-energy electrons could be eliminated. The exact location and the cause of the negative blocking voltage are not presently known, but will be investigated during future commissioning measurements. Since the possible blocking potential is limited to a small portion of the beamline, however, it will not reduce the final kinetic energy of electrons that pass this region and enter the main spectrometer. At the same time, since it is very small (\SI{<100}{\volt}) compared to the \betael{} energy in the endpoint region (about \SI{18.6}{\kilo\electronvolt}), only \betaels{} with a very small longitudinal momentum component will be blocked. For example, if the blocking potential is located in the center of the downstream pre-spectrometer magnet, blocked endpoint electrons have pitch angles greater than \SI{85.8}{\degree}. \betaels{} with such large pitch angles are already excluded from the KATRIN acceptance via magnetic reflection (\sref{apparatus:sds}), so this possible blocking potential will have no influence on the neutrino-mass measurements.

\subsection{Backgrounds with electron source}
\label{sec:FirstLightBackground}
In the standard KATRIN operational mode, positive ions will be created inside the beamline and the spectrometer section due to ionization of residual gas molecules by the large flux of about \num{e10} \betaels{} per second. These ions are non-adiabatically accelerated by the retarding potential inside the main spectrometer (\sref{apparatus:sds}) and ionize residual gas molecules in a region where \betaels{} are themselves reduced to very low energy. Secondary electrons produced in this process will be magnetically guided to the FPD, where they will be detected with the same energy as the surviving \betaels{}.

In order to investigate this background mechanism, the UV irradiation system at the rear wall (\sref{sec:rw_pe}) was reconfigured for an increased electron rate, approximately \num{1.6+-0.5e8}~electrons per second at an energy of \SI{110}{\electronvolt}. The main-spectrometer background was then measured at a nominal retarding potential of \SI{-18.6}{\kilo\volt}. Without any UV irradiation, the reference background was measured at \SI{599+-4}{\milli cps}, with both the pre-spectrometer and its downstream ring electrode (\sref{apparatus:sds}) set to ground potential; with maximum UV irradiation, the background increased by a factor of four, to about \SI{2.5}{cps}. This background was reduced significantly by applying \SI{+300}{\volt} to the ring electrode, but it was still elevated at \SI{705}{\milli cps}; we attribute the remaining excess to ionization processes downstream of the ring electrode. Operating the pre-spectrometer at a negative potential, \SI{-300}{\volt}, was sufficient to return the background rate to the reference value, with or without the additional ring electrode; this configuration blocked the low-energy electron flux and non-adiabatically accelerated positive ions out of the flux tube, preventing them from entering the main spectrometer. 

\begin{table}[tbp]
	\caption{\label{BackgroundsWithElectronSourceTable}Background rate for different blocking configurations for ions and electrons, at the maximum achieved photo\-electron flux. The reference background without UV irradiation of the rear wall was \SI{599+-4}{\milli cps}.}
	\begin{center}
		\begin{tabular}{lrr}
			\hline
			&Pre-spec.\ grounded & Pre-spec.\ about \SI{-300}{\volt}\\
			\hline
			Ring-electrode ion blocking off& \SI{2530+-90}{\milli cps}& \SI{590+-20}{\milli cps}\\
			Ring-electrode ion blocking \SI{300}{\volt}& \SI{705+-17}{\milli cps}&\SI{589+-16}{\milli cps}\\
			\hline
		\end{tabular}
	\end{center}
\end{table}

From these measurement results, summarized in \tref{BackgroundsWithElectronSourceTable}, we conclude that during the neutrino-mass measurements the \betaels{} will produce some background by ionization. However, due to the different operating conditions for the neutrino-mass run (i.e., lower pressure, smaller ionization cross section due to larger electron energy, and reduced electron reflection from the rear wall), this background component is expected to be of order of \SI{0.01}{\milli cps}  with a grounded pre-spectrometer. This is comparable with earlier results~\cite{Prall2012}.

\subsection{Detection of deuterium ions}
\label{sec:ion_detection}
Tritium ions, created in the WGTS by \betadec{} and ionization interactions, could cause significant background if they enter the spectrometers. During tritium running, these ions will be blocked and removed from the flux tube~\cite{klein:2017}. The ELIOTT ion source (\sref{sec:eliott}) was configured to produce deuterium ions in order to test the basic functions of these planned ion-blocking and ion-removal methods. The ion pencil beam was detected via ionization of residual gas in the spectrometers (\sref{apparatus:sds}). When the positive deuterium ions encountered the negative potential of the spectrometers, they were accelerated non-adiabatically toward the vessel walls. On their way across the unbaked spectrometers, they scattered on the residual gas, dominated by \h2o{}. The secondary electrons were then counted with the FPD as a linear proxy for the ion flux that entered the spectrometers. Observation of relative rate changes allowed assessment of the ion-blocking and ion-removal instruments in the beamline.

The ion detection efficiency was determined as 
\begin{equation}
	\mathcal{E_{\mathrm{ion}}} = \frac{R_\mathrm{FPD}}{J_{\mathrm{ion}}},
\end{equation}

\noindent where $R_\mathrm{FPD}$ is the detected FPD rate and $J_{\mathrm{ion}}$ is the ion current entering the main spectrometer. With both spectrometers grounded, $J_{\mathrm{ion}}$ was measured using a Faraday cup, instrumented with PULCINELLA, in the FPD system (\sref{sec:fpd}). With the main spectrometer at \SI{-18.6}{\kilo\volt}, a residual \h2o{} pressure of order \SI{e-9}{\milli\bar}, and with ions of about \SI{30}{\electronvolt} of kinetic energy, we found $\mathcal{E}=$~\SI{1.18+-0.05e-4}{FPD} counts per ion, quoting only statistical uncertainty. 

\subsection{Test of active methods for ion blocking and removal}
\label{sec:ionblocking}
The deuterium ions were blocked successfully with positive potential on the four beamline ring electrodes. Suppression factors of at least $\sim$\num{155} (transport-section electrodes), \num{43} and \num{130} (pre-spectrometer electrodes) were measured, limited primarily by the ion flux and the measurement background.

With the ring electrodes off, the ion pencil beam produced a hot spot on the FPD pixels, which could be moved with an $\vec E\times \vec B$ drift induced by the dipole electrodes in the transport section (\sref{apparatus:transport}). The hotspot dissolved quickly and the complex pattern of the FPD pixels does not allow the drift distance of the pencil beam to be determined for a specific dipole electrode voltage. In agreement with analytical estimates, the \SI{2}{\electronvolt} ions were completely removed when one approximately \SI{1}{\metre} long dipole electrode was set to about \SI{-360}{\volt} in a magnetic field of approximately \SI{1}{\tesla}.

\begin{figure}[tbp]
	\centering
	\includegraphics[width=0.8\textwidth]{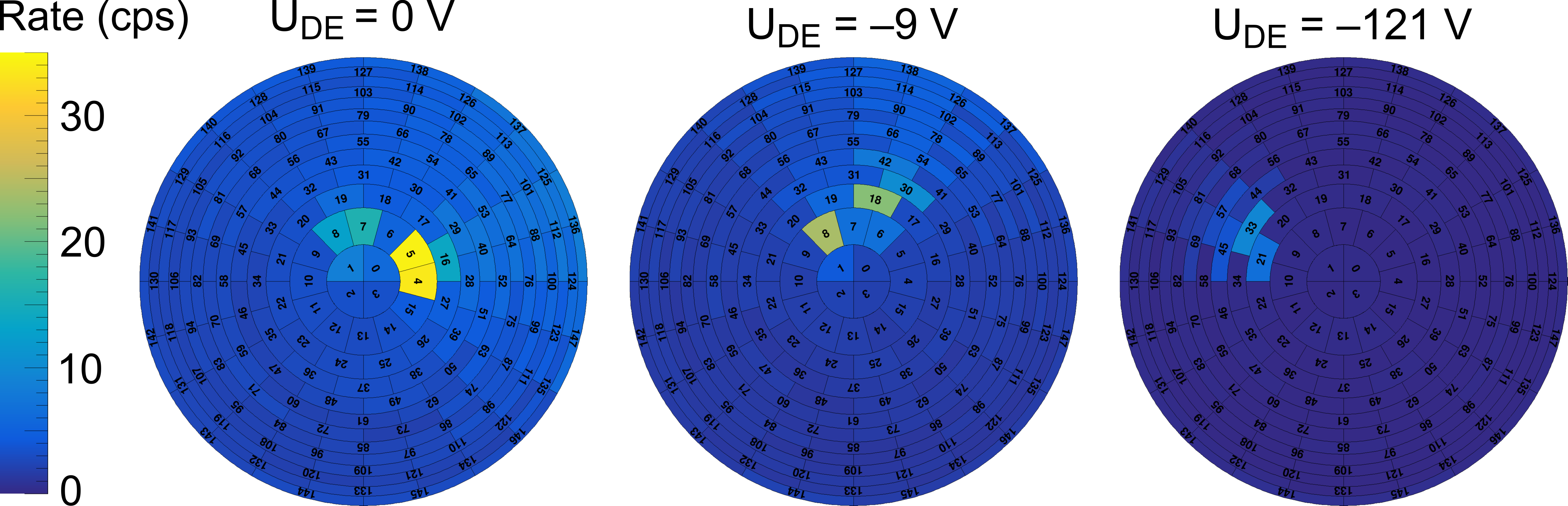}
	\caption{Pixel view of the secondary-electron rate at the FPD as a DPS dipole electrode was used to drift the ion beam. The ion beam was shifted to the left with increasing electrode voltage. The evolution of specific features needs more detailed investigation.}
	\label{pic.ionhotspot}
\end{figure}

The negative potential of the pre-spectrometer can also reduce the ion flux into the main spectrometer by accelerating the ions non-adiabatically toward the vessel walls. When the pre-spectrometer was set to \SI{-1}{\kilo\volt}, the flux of \SI{2}{\electronvolt} ions was reduced by a factor of about \num{741}. 

\subsection{Beamline blocking of thermal ions}
\label{sec.unbeabsichtigteIonenblockierung}
Ions with thermal energies were blocked in the beamline. The blocking was discovered in the ion energy spectrum which was measured by gradually increasing the positive blocking potential provided by a single DPS dipole electrode in monopole mode. The differential spectrum shows a peak at the ELIOTT electrode voltage, but no Maxwell-Boltzmann peak at energies \SI{<1}{\electronvolt} as would have been expected due to thermalization in the dense deuterium gas at room temperature in ELIOTT. However, a sharp peak, separated from background by one order of magnitude, appeared when the ions were given an energy offset of about \SI{2}{\electronvolt} by setting the ELIOTT hull to \SI{+4}{\volt} and the rear wall to \SI{+2}{\volt}. Clearly, ions with energies $\lesssim$\SI{1}{\electronvolt} were blocked in the beamline by a small positive potential. The most likely explanation are work-function differences of order \SI{100}{\milli\electronvolt} along the beamline, as expected due to different beamline materials and adsorption of residual gas on the surface. An alternative explanation is a local positive charge on the insulated cables within the DPS flux tube, charged up to \SI{120}{\volt} by the previously drifted ion beam and producing a cut-off coincidentally on the order of the expected work-function effect.

Thermal-ion blocking during the neutrino-mass measurements would lead to an accumulation of ions inside the WGTS, and the increased plasma density could affect the \betael{} energy via time-varying electric fields during plasma instabilities. As a result of these measurements, a fourth dipole electrode was inserted into the first DPS beam-tube element to remove stored ions as close as possible to the WGTS. This electrode will be tested in future commissioning experiments.

\section{Gaseous \kr{} campaign}
\label{sec:GKrS-campaign}

Gaseous \kr{} mixed into the tritium gas in the WGTS can be used to determine plasma properties of the tritium source with high accuracy. For instance, modifications of the well-known conversion-electron line shapes can reveal charging effects in the source gas that would affect the measurement of the tritium \betaspec{}. The gaseous \kr{} campaign described in this section serves as a first reference run of \kr{} spectroscopy without any additional carrier gas inside the source section, in order to characterize the KATRIN apparatus with respect to high-voltage operation and temperature-dependent source properties. A dedicated krypton generator (\sref{sec:GKrS}) was installed at the front pumping chamber of the WGTS (\sref{sec:wgts}), injecting \kr{} into a beam tube maintained at \SI{100}{\kelvin}; this temperature is well above krypton freezeout at pressures below \SI{100}{\milli\bar}~\cite{leming:1970}. The \SI{4.5}{\kelvin} CPS beam-tube surface (\sref{apparatus:transport}) maintained an in-beamline pressure of less than \SI{1e-8}{\milli\bar}, so that additional pumping with TMPs was not necessary. To monitor source activity, a single BIXS system (\sref{apparatus:BIXS}) was connected at the rear of the source tube. In addition, the FBM (\sref{apparatus:FBM}) was operated inside the CPS, nondestructively measuring electron signals and magnetic fields throughout the measurement cycle. In order to differentiate signals originating from the WGTS and DPS, the initial energy of the decay electrons in the DPS (\sref{apparatus:transport}) was decreased with four dipole electrodes at \SI{+350}{\volt}. Ion blocking was realized by holding three ring electrodes, all upstream of the pre-spectrometer, at a constant potential of \SI{200}{\volt}. The vacuum-baked main spectrometer (\sref{apparatus:sds}), at an ultimate pressure on the order of \SI{e-11}{\milli\bar}, was used to analyze the \kr{} spectrum; the unbaked pre-spectrometer was held at ground potential. The monitor spectrometer (\sref{sec:mos}) was operated in parallel.  

Below, we report on early results from this commissioning campaign. \secref{subsec:GKrS:first-electrons} describes the characteristics of the electrons observed in three beamline detectors: the FPD, the FBM, and the BIXS system. \secref{subsec:GKrS:activity-as-seen-in-fpd-and-kr-distribution} reports on a study of the observed \kr{} activity and its distribution within the beamline, based on FPD data. Qualitative information about the line shapes, including sample integrated spectra, is given in \sref{subsec:GKrS:qualitative-information-about-line-shapes}.  \secref{subsec:GKrS:system-stability} demonstrates the achieved system stability along the entire beamline.

\subsection{First electrons}
\label{subsec:GKrS:first-electrons}
Throughout the gaseous \kr{} campaign, three detector systems were operational along the KATRIN beamline. The FPD (\sref{sec:fpd}) observed electron spectra produced by energy scans with the main spectrometer (\sref{apparatus:sds}), and the two monitoring systems (BIXS, \sref{apparatus:BIXS}, and FBM, \sref{apparatus:FBM}) observed electron and X-ray spectra originating directly from the WGTS (\sref{sec:wgts}). These overlapping detection methods provided an overall picture of the first \kr{} electrons within KATRIN, and allowed studies of the source-activity stability during the measurement \kr{} campaign (\sref{subsubsec:GKrS:first-electrons:count-rate-stability}).

\begin{figure}[tbp]
	\centering
	\includegraphics[width=0.7\textwidth]{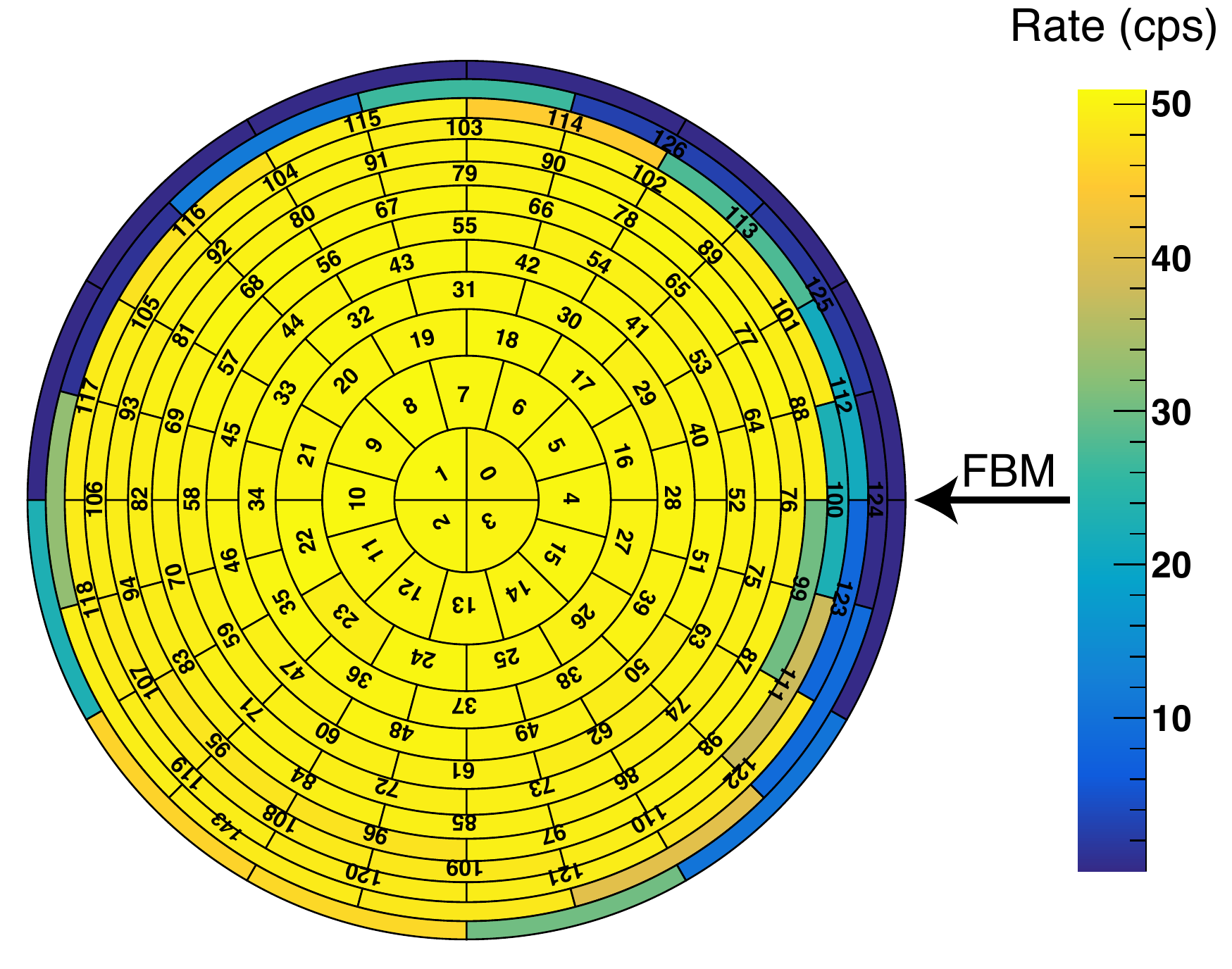}
	\caption{FPD pixel map of electrons originating from the GKrS with region-of-interest cut for the M$_1$-32 line (expected energy \SI{31.9}{\kilo\electronvolt}). Electrons of other conversion lines are also included conditioned by the cut range. The color scale represents the rate on the FPD; the arrow shows the FBM direction of motion when entering the beamline. A homogeneous irradiation of almost all detector pixels was observed.}
	\label{fig:FPD_pixel_map_krypton}
\end{figure}

The FPD detected electrons transmitted through the main spectrometer. For a given spectrometer setting, an average count rate was determined for each FPD pixel. For this analysis, a region of interest was defined for the measured electron energy $E_e$ and charge $q$, according to \eref{eq:E0_ROI}:

\begin{equation}
\label{eq:E0_ROI}
	E_i + q(U_{\mathrm{PAE}} + U_{\mathrm{BIAS}}) - \SI{3}{\kilo\electronvolt} \leq E_e \leq E_i + q(U_{\mathrm{PAE}} + U_{\mathrm{BIAS}}) + \SI{2}{\kilo\electronvolt},
\end{equation}

\noindent where $E_i$ is the expected conversion-electron energy for the scanned line, $U_{\mathrm{PAE}} = \SI{10}{\kilo\volt}$ the post-acceleration voltage (\sref{sec:fpd}) and $U_{\mathrm{BIAS}} = \SI{0.12}{\kilo\volt}$ the bias voltage of the FPD. The asymmetric range accounts for electron energy loss in the insensitive dead layer. A pixel map of the measured rates in a typical run is shown in \fref{fig:FPD_pixel_map_krypton}. A homogeneous irradiation of almost all detector pixels was observed. The outer two pixel rings show a decreased rate indicating the flux tube boundary. A slight misalignment of the flux tube is noticeable at the bottom left border of the FPD image. 

The FBM measured electron spectra in the vicinity of the outer rim of the flux tube in the CPS, before energy filtering in the main spectrometer. During the gaseous \kr{} campaign this location mapped to FPD pixel \num{100}, resulting in a slight shadowing on the FPD as shown in \fref{fig:FPD_pixel_map_krypton}. The temperature and magnetic field were also measured during the gaseous \kr{} campaign at this location in the beam. For each analysis run, an average count rate was calculated from the krypton spectra measured by the \pin{} diodes after applying an energy threshold of \SI{30}{\kilo\electronvolt}. In \sref{subsubsec:GKrS:first-electrons:count-rate-stability}, we analyze the stability of this count rate in comparison to FPD and BIXS data.

\begin{figure}[tbp]
	\centering
	\includegraphics[width=1.0\textwidth]{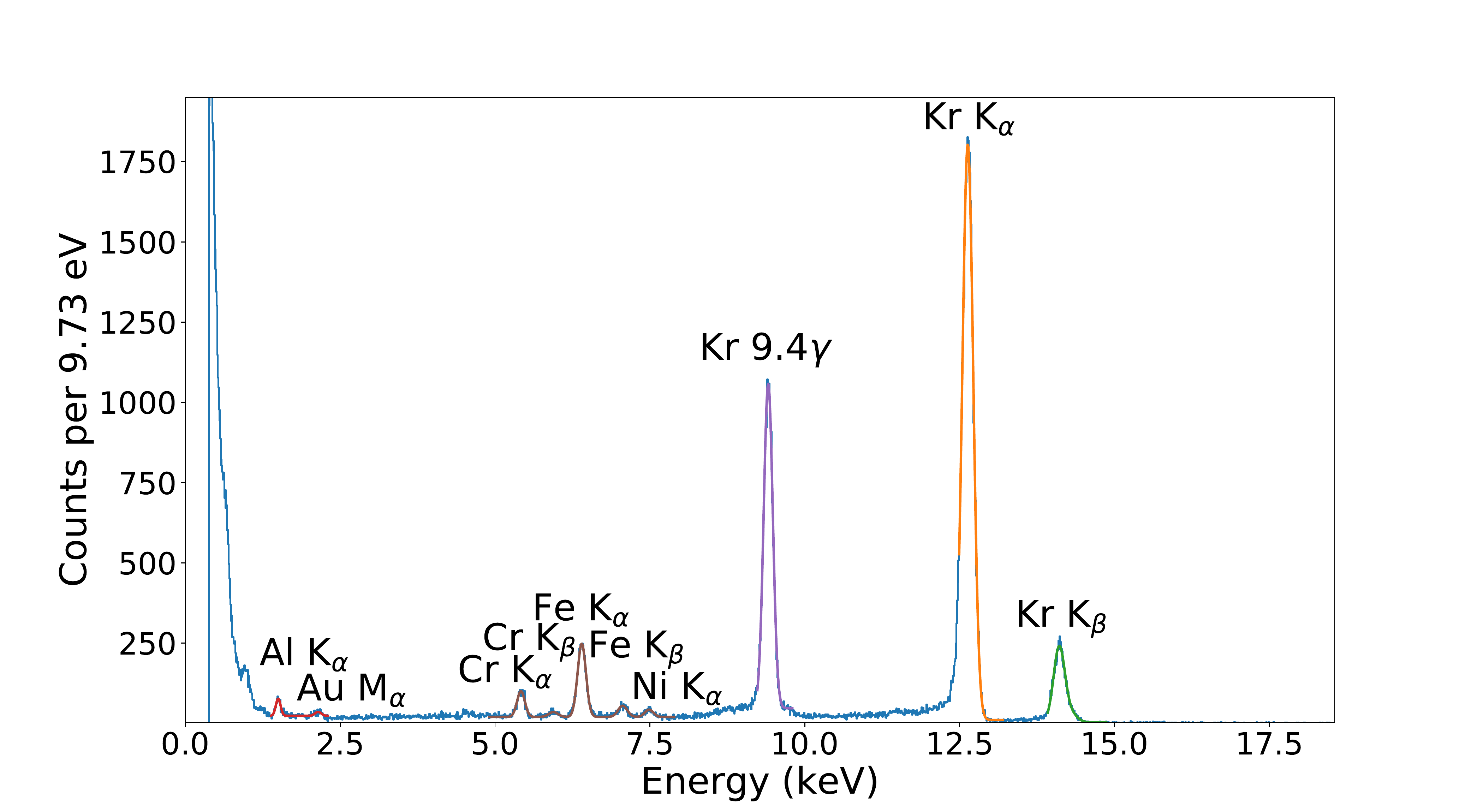}
	\caption{Accumulated X-ray spectrum from three days of measurement. The detector was calibrated \emph{in situ} by fitting the \kr{}~K$_{\alpha}$, the \SI{9.4}{\kilo\electronvolt} $\upgamma$, and the Fe~K$_{\alpha}$ peaks. For the three most prominent peaks the fits were performed asymmetrically because of their low-energy Compton tails. For energies above \SI{15}{\kilo\electronvolt} only background was measured.}
	\label{fig:BIXS_accumulated_spectrum}
\end{figure}

The BIXS system is intended to measure X-ray events produced by bremsstrahlung of \betaels{}. During the gaseous krypton measurements this was not the dominant X-ray production mechanism; as seen in \fref{fig:BIXS_accumulated_spectrum}, the continuum was suppressed in the spectrum, which was dominated by characteristic lines. One of these originated from the \SI{9.4}{\kilo\electronvolt} $\upgamma$-decay. The X-ray peaks with the highest count rates were associated either with \kr{} or with iron, nickel or chromium, the most prominent components of the stainless steel comprising the WGTS beam tube. Two low-amplitude peaks originated from the gold and aluminum of the electric contacts on the silicon chip. During each analysis run, an integrated count rate was obtained by summing over all data channels.

\subsection{Distribution of \kr{} gas and source activity}
\label{subsec:GKrS:activity-as-seen-in-fpd-and-kr-distribution}
We checked our understanding of the source activity by estimating the FPD count rate due to the \SI{32}{\kilo\electronvolt} transition, starting with the \SI{1.0}{\giga\becquerel} activity of the \textsuperscript{83}Rb source and applying known reduction factors. The metastable state \kr{} is produced in this decay with a probability of \SI{74}{\percent}~\cite{Mccutchan2015}. The \textsuperscript{83}Rb was embedded in zeolite beads (\sref{sec:GKrS}), and between \SI{80}{\percent} and \SI{90}{\percent} of the \kr{} diffused out as determined in earlier tests~\cite{Sentkerestiova2017}. The FPD electron-detection efficiency was taken as 
$\SI{95.0}{\percent} \pm \SI{1.8}{\percent}_\text{\,stat.} \pm \SI{2.2}{\percent}_\text{\,sys.}$~\cite{Amsbaugh2015}. The branching ratio of \kr{} decay to the M-32 and N-32 lines was taken as \SI{11}{\percent}~\cite{Mccutchan2015}. 

\begin{figure}[tbp]
	\centering
	\includegraphics[width=1.0\textwidth]{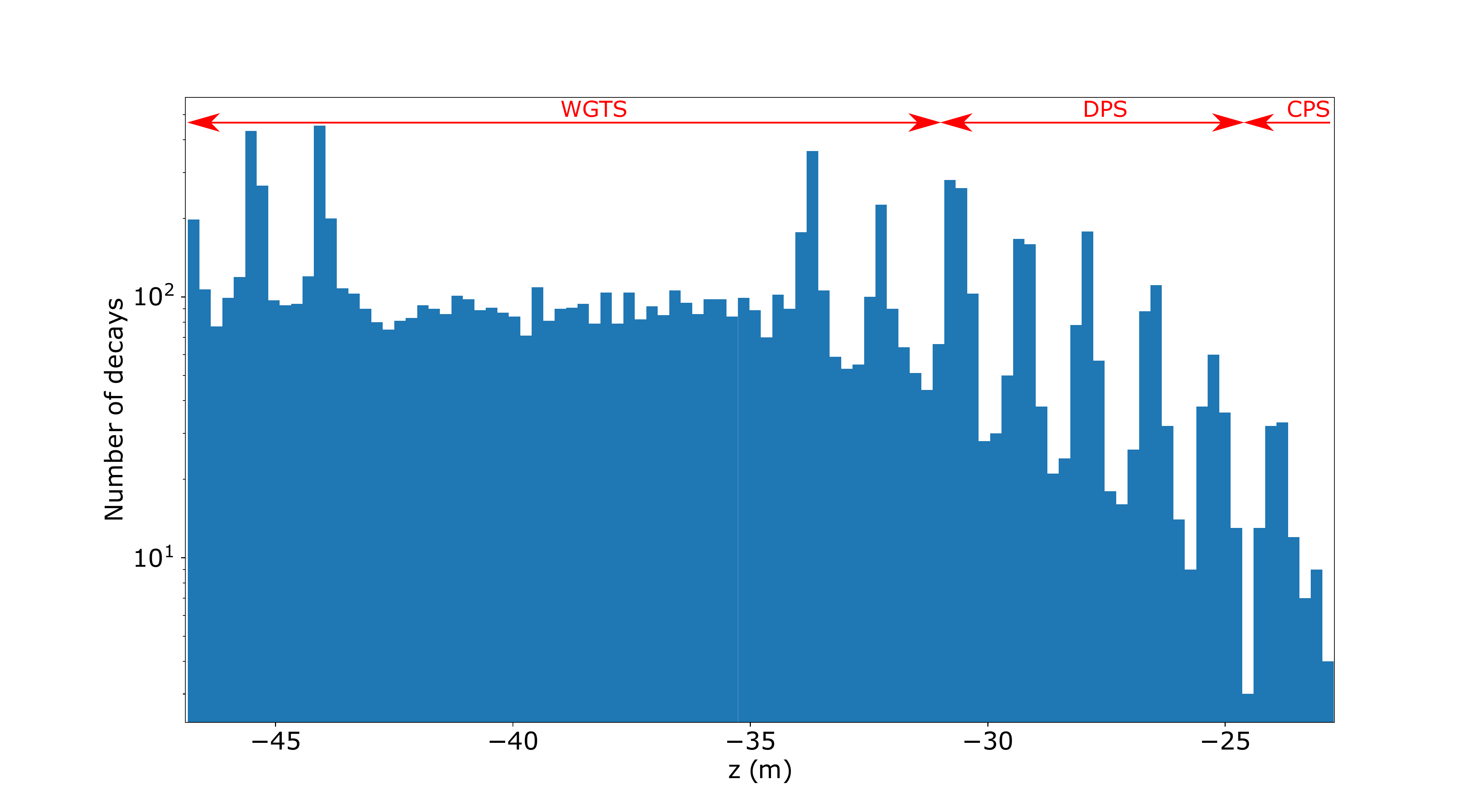}
	\caption{Lateral distribution of \kr{} decays inside the magnetic flux tube, as simulated with MolFlow+. The positions along the beam tube axis $z$ are given in KATRIN coordinates, with the origin in the center of the main spectrometer. Towards the CPS, which was the only active pump in this configuration, the number of decays decreased. The significant peaks were located at the pumping port positions of the beam line. Here the magnetic flux tube widened, increasing the number of decays within -- although the excess is ultimately removed from the acceptance by the magnetic-mirror effect.}
	\label{fig:krypton-distribution}
\end{figure}

Finally, there was an additional reduction associated with the distribution of the \kr{} inside the beam-tube volume: only those electrons created inside the magnetic flux tube and emitted within the maximum accepted pitch angle were guided to the detector. To determine these factors, we performed simulations with a modified version of MolFlow+~\cite{Molflow2016,Drexlin2017} that enables the tracking of radioactive atoms in molecular flow and the determination of decay positions. The simulation included a detailed model of the transport section along with the temperature $T$ of each element. The mean sojourn time of atoms on a surface,

\begin{equation}
\label{eq:sojtime}
\tau_{\mathrm{des}}=\tau_0 e^{E_{\mathrm{B}}/(RT)},
\end{equation}

\noindent was implemented with literature values for the binding energies $E_B$ for gold and stainless-steel surfaces~\cite{Troy1971, Wischlitzki1975}. The universal gas constant $R$, the temperature $T$ and the period of the sticking particle's oscillation perpendicular to the surface $\tau_0$ = \SI{1e-13}{\second}~\cite{Haefer1981} are well known. The sticking coefficients of \kr{} cryosorption on stainless steel and on gold, which give the probabilities that \kr{} sticks to these surfaces, were set to \num{1} since no literature values exist. Of the simulated \kr{} atoms only \SI{0.26}{\percent} were found to decay in the magnetic flux tube. For those decays, the distribution along the beam-tube axis is plotted in \fref{fig:krypton-distribution}. The simulated average probability of a conversion electron being emitted within the maximum accepted pitch angle is 12.9\%.

Applying these reduction factors to the \kr{} activity of \SI{1.0}{\giga\becquerel}, we reached an expected count rate of approximately \SI{22.0}{\kilo cps} for the M-32 and N-32 lines. A measurement of electrons from all such lines yielded a count rate of ($23.566 \pm 0.013_{stat.}$)~kcps, a satisfactory level of agreement. In addition to the unknown sticking coefficients, the analysis does not account for reductions in the effective flux-tube volume due to misalignment. 

During these measurements, there was no pumping in the WGTS or DPS, so we also investigated how the \kr{} decays were distributed between these sections. For one measurement phase, both half-shells of the electric dipole elements in the DPS were set to \SI{350}{\volt}, shifting the electrons coming from this region to lower energies. Later, to exclusively measure the decays from the DPS, the dipole elements were set to \SI{-350}{\volt}. Simulations, performed with the Kassiopeia software package~\cite{Furse2017}, showed that, in those portions of the flux tube inside the DPS pumping ports, the electric potential went down to nearly \SI{0}{\volt}. Therefore, it is only possible to compare count rate ratios originating from different starting potentials, instead of completely separating decays from the WGTS and the DPS. We note that, with differential pumping operational during neutrino-mass running, the tritium decay rate from the DPS will be negligible.

Combining the MolFlow+ and the Kassiopeia simulations, a ($15.4 \pm 0.6_{stat}$) times higher count rate was expected to come from regions of the beam tube with zero electric potential than from regions with potential near \SI{-350}{\volt}. In comparative measurements of the L$_3$-32 line with positive and negative polarity of the dipole voltage (\SI{\pm350}{\volt}), this factor was found to be $14.0 \pm 2.4_{stat.}$, in good agreement with the simulation. 
The residual discrepancy is likely due to the relatively low statistics of the simulation, and to the unknown \kr{} cryosorption sticking coefficients.

\subsection{Line shapes}
\label{subsec:GKrS:qualitative-information-about-line-shapes}

The complete \kr{} conversion-electron spectrum was measured using the main spectrometer with an energy resolution (\eref{eq:mac-e}) of $\Delta E / E =$ \SI{2.7e-4}{\tesla}/\SI{4.2}{\tesla}, corresponding to about $\Delta E =$ \SI{1.15}{\electronvolt} for the K-32 line. The retarding-potential step size and measurement time per voltage step were adjusted for each line based on its natural width and relative intensity.

\begin{figure}[tbp]
	\centering
	\includegraphics[width=0.8\textwidth]{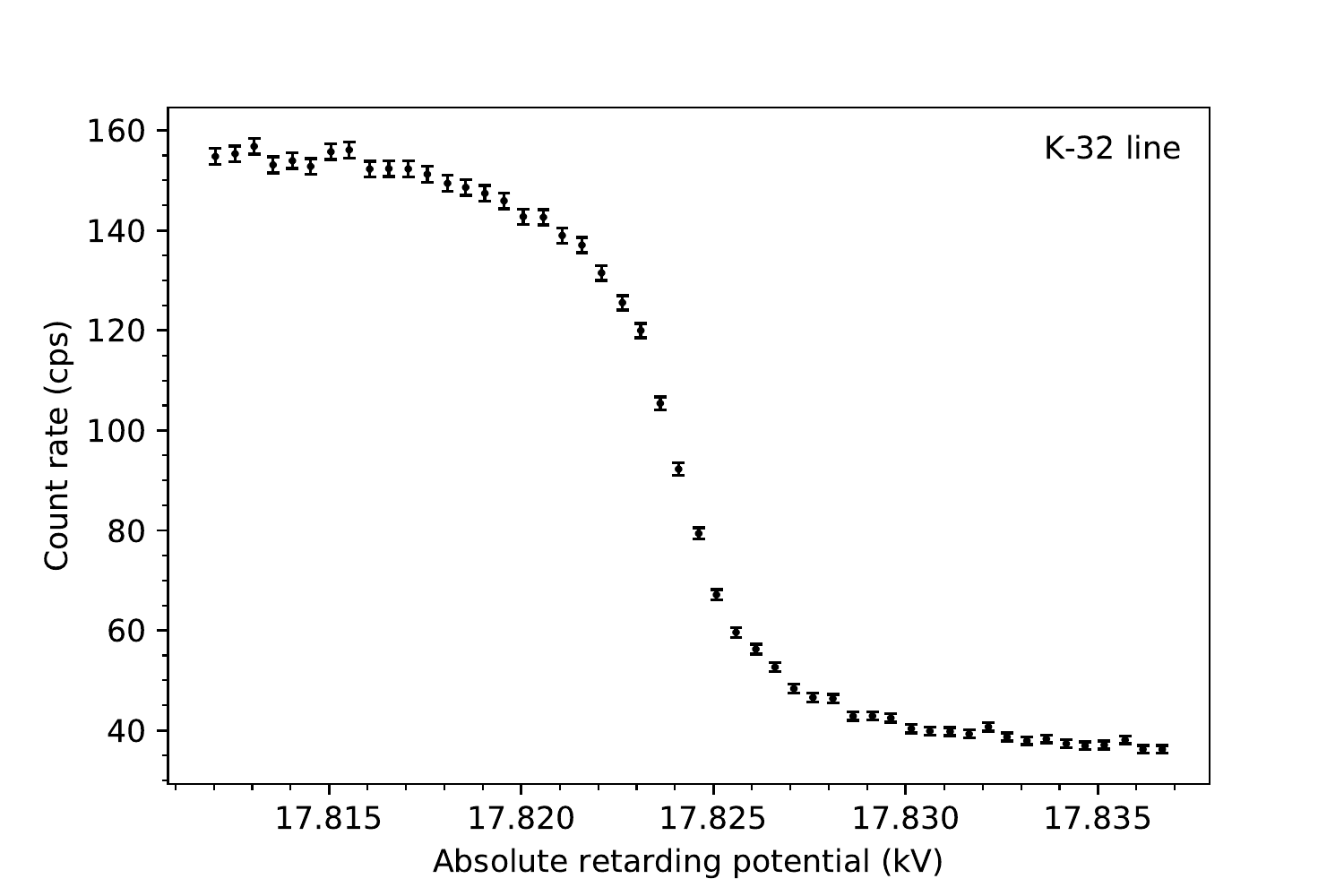}
	\caption{Integral spectrum of the K-32 line obtained with the gaseous \kr{} source. The horizontal axis gives the absolute value of the retarding-potential reading, corrected for the mean potential depression at the analyzing plane as seen by the bullseye pixels (\sref{apparatus:sds}). The retarding-potential step size was \SI{0.5}{\volt} and the measurement time at each step was \SI{60}{\second}. The electron counts were summed over the four bullseye FPD pixels.}
	\label{fig:GKrS_K-32}
\end{figure}

The integral spectrum of the K-32 line is shown in \fref{fig:GKrS_K-32}. The typical transmission function of a MAC-E spectrometer (\eref{eq:transmission}) assumes a mono-energetic source. In contrast, the K-32 line features a relatively large natural line width (approximately \SI{2.7}{\electronvolt}~\cite{venos:2018}). This results in a smearing of the otherwise sharply defined transmission edges (compare \fref{fig:transmission}), which is especially visible towards the edge of the spectral shape where the slowly decreasing Lorentzian tails of the K-32 line come into effect. Hence, the measured spectrum shows the shape expected from a MAC-E filter observing an isotropic electron source~\cite{Picard1992b} and therefore demonstrates the spectrometer's high energy resolution. 
With a full and careful treatment of the spectrometer transmission function, as well as energy losses in transport, a differential spectrum may be extracted from this integral measurement. A forthcoming publication~\cite{Slezak2018} will present differential spectra measured during the GKrS campaign.

\begin{figure}[tbp]
	\centering
	\includegraphics[width=0.8\textwidth]{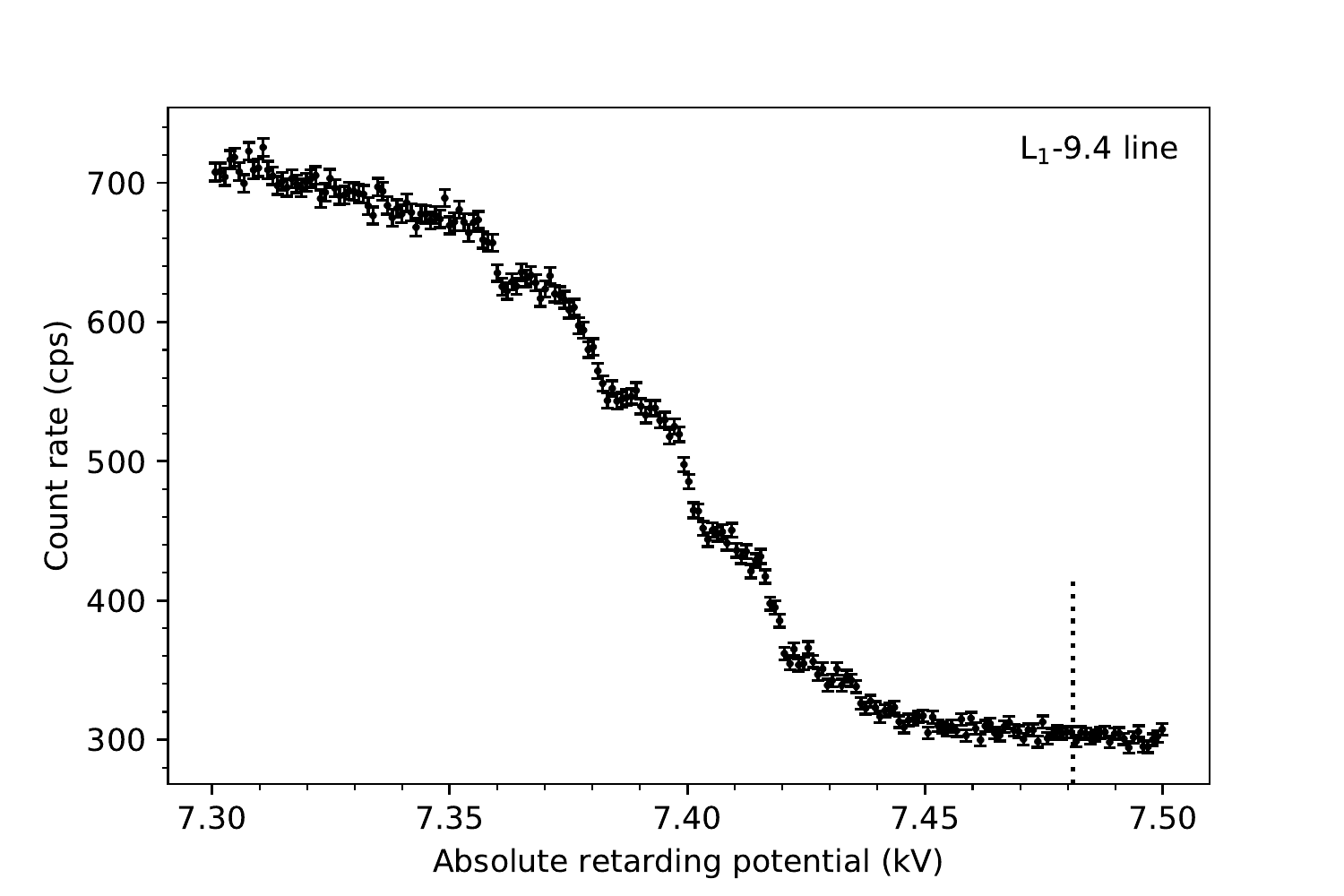}
	\caption{The multiplet structure of the \SI{9.4}{\kilo\electronvolt} conversion electrons obtained with the gaseous \kr{} source. The retarding-potential step size was \SI{1}{\volt} and the measurement time at each step was \SI{18}{\second}. The electron counts were summed over the four bullseye FPD pixels. An offset corresponding to the mean potential depression at the analyzing plane (\sref{apparatus:sds}), as seen by the inner pixels, was applied to the retarding potential reading. The dotted vertical line indicates the energy of the high-intensity L$_{1}$-9.4 transition in a neutral krypton atom.}
	\label{fig:GKrS_Hedgehog-L}
\end{figure}

A notable feature was observed when measuring \SI{9.4}{\kilo\electronvolt} conversion electrons. The decay of \kr{} to its ground state involves a cascade of two transitions. Multiple Auger electrons are emitted in the de-excitation of the atomic shell after the first internal conversion, and the krypton atom is thus left in a possibly highly ionized state. Due to the low density of the krypton gas in our source and to the short half-life of the remaining nuclear excited state (about \SI{150}{\nano\second}~\cite{Mccutchan2015}), there was not enough time for the atom to neutralize before the second nuclear transition occurred. Thus, the corresponding conversion electrons had different kinetic energies, depending on the ionization state of the krypton atom. A multiplet of lines thus emerged in the gaseous \kr{} spectrum, as previously described by Decman and Stoeffl~\cite{Decman1990}. \figref{fig:GKrS_Hedgehog-L} shows part of the integral spectrum of the multiplet structure, corresponding to the L-shell electrons and measured with the main spectrometer.

\subsection{System stability}
\label{subsec:GKrS:system-stability}
The system stability during the gaseous \kr{} campaign governed the reliability of analysis and results of the measured data. Long-term stability analyses were performed on \kr{} activity as measured in the generator (\sref{subsubsec:GKrS:system-stability:activity-stability}) and in the detectors (\sref{subsubsec:GKrS:first-electrons:count-rate-stability}), and on the \kr{} line position and width (\sref{subsubsec:GKrS:system-stability:line-stability-in-fpd}). The results of long-term stability analyses for other device parameters (\sref{subsubsec:GKrS:system-stability:slow-control-stability}), high-frequency stability analysis on the high-voltage system (\sref{subsec:GKrS:high-voltage-ripple}), and continuous observation of monitor-spectrometer data (\sref{subsubsec:GKrS:system-stability:monitor-spectrometer-data}) may also be applied to the CKrS measurement campaign (\sref{sec:CKrS-campaign}).

\subsubsection{Activity stability in GKrS generator}
\label{subsubsec:GKrS:system-stability:activity-stability}
The production of gaseous \kr{} in the generator followed the \textsuperscript{83}Rb decay curve with its half-life of \SI{86.2}{\day}~\cite{Mccutchan2015}, so the \kr{} activity was expected to decrease roughly \SI{1}{\percent} per day, assuming stable emanation as previously demonstrated~\cite{Venos2014}. The source activity was measured at the generator using a $\upgamma$-ray spectroscopy method~\cite{Venos2014,Sentkerestiova2017}, for which an Amptek X-123 X-ray spectrometer with a Si drift detector was installed with a view of the generator chamber. The spectrometer detected K X-rays and \SI{32.2}{\kilo\electronvolt} $\upgamma$s from decay of \kr{} gas within the generator.

At the beginning of the measurement campaign, this spectrometer measured $\pm$\SI{0.8}{\percent} fluctuations in gaseous \kr{} activity over two hours. Immediately after the valve was opened between the generator and the WGTS, the amount of gaseous \kr{} inside the generator dropped precipitously, resulting in a K X-ray rate reduction by a factor of \num{152}. With more limited statistics, the observed activity fluctuations were at the $\pm$\SI{8}{\percent} level. In both cases, the observed fluctuations were approximately a factor of two higher than expected from counting statistics. Beamline detectors --- the BIXS, FBM, and FPD --- were used to monitor the overall activity stability with the valve open, as reported in \sref{subsubsec:GKrS:first-electrons:count-rate-stability}. The FPD also established a $\pm$\SI{0.8}{\percent} activity fluctuation level in \kr{} activity.

\subsubsection{Count-rate stability}
\label{subsubsec:GKrS:first-electrons:count-rate-stability}
\begin{figure}[tbp]
	\centering
	\includegraphics[width=0.8\textwidth]{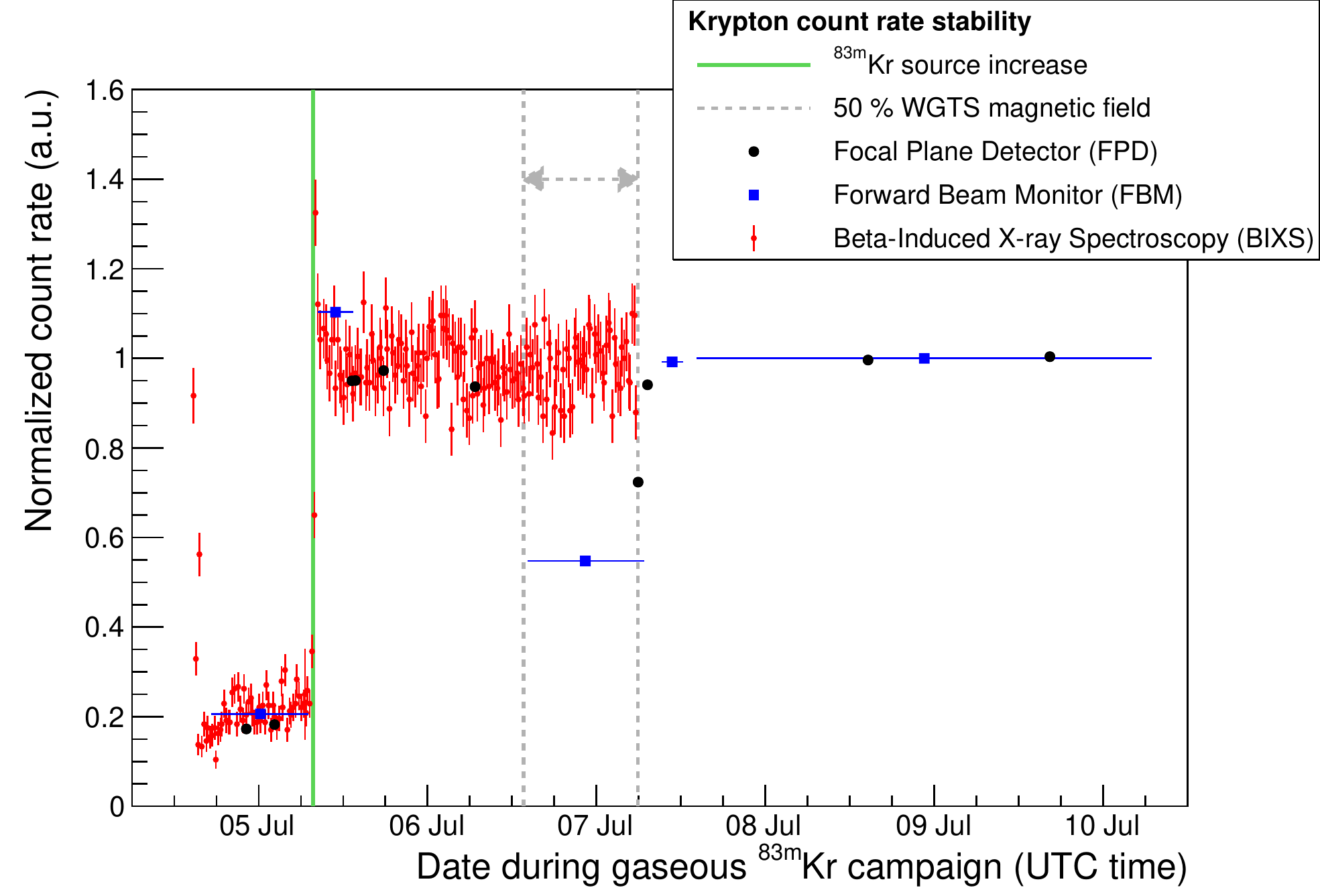}
	\caption{Normalized relative count rates from three KATRIN  detector systems during the gaseous \kr{} campaign. FPD data, at a \SI{0.27}{\milli\tesla} setting, are shown in black circles; FBM data are shown in blue squares; and BIXS data are shown in red dots. The data from all three detectors are normalized to one in their respective stable regions. Vertical lines indicate hardware changes during the campaign. All three detectors observed approximately a fivefold increase in count rate when the \kr{} source intensity was increased (green line). The FPD and FBM observed a decrease in count rate (approximately \SI{20}{\percent} and \SI{50}{\percent}, respectively) when the WGTS had a lowered magnetic field (gray lines), while BIXS was insensitive to the WGTS magnetic field. Statistical errors in FPD and FBM data are too small to be visible. BIXS data-taking was stopped on 7 July.}
	\label{fig:FPD_FBM_BIX_count_rates}
\end{figure}

The relative count rates from the three KATRIN detector systems are shown in \fref{fig:FPD_FBM_BIX_count_rates}. During the gaseous \kr{} campaign two hardware changes were made, modifying the electron flux intensity and distribution: a one-time increase of the amount of \kr{} introduced from the generator to the source, and a 16-hour period in which the magnetic field in the WGTS was lowered to \SI{50}{\percent} of its nominal value. The times of these hardware changes are indicated by the vertical lines in \fref{fig:FPD_FBM_BIX_count_rates}. 

When the intensity of the \kr{} source was increased, all three detector systems observed an approximately fivefold increase in count rate. The BIXS system is also sensitive to rate changes happening on a time scale of \SI{1000}{\second}, due to manual adjustments of the regulation valve at the \kr{} generator. The FPD and FBM observed lower count rates during the overnight measurements with lower magnetic field in the WGTS. The lower count rate is due to a combination of the lower acceptance angle of electrons from the source and the smaller, more focused electron flux tube. This second effect is radially dependent, slightly increasing the count rate towards the center of the flux tube, and strongly decreasing the count rate towards the outside of the flux tube. The FBM observed an approximate \SI{50}{\percent} decrease in count rate at the outer edge (equivalent to pixel \num{100} on the FPD; see \fref{fig:FPD_pixel_map_krypton}), while the FPD observed an approximate \SI{20}{\percent} overall decrease. As expected, the BIXS system, which imaged the full WGTS flux tube by measuring X-rays, observed no effect from the lower magnetic field settings in the WGTS.

\subsubsection{Line stability in FPD}
\label{subsubsec:GKrS:system-stability:line-stability-in-fpd}
The FPD (\sref{sec:fpd}) measured a differential spectrum of electrons that passed the energy threshold set by the main spectrometer, with an intrinsic energy resolution on the order of a few keV (FWHM). Higher-energy \kr{} lines could thus be reconstructed by the FPD from main-spectrometer scans of lower-energy lines. We studied the positions and widths of these reconstructed lines to search for instabilities in detector performance and the experimental energy scale.

For this stability analysis, eight runs scanning the M$_{1}$-32 line were selected, spanning both the GKrS and CKrS campaigns. A fit to the peak's position revealed that though there is some fluctuation at the \SI{1}{\percent} level, there was no cohesive pattern or trend. These fluctuations are not large enough to affect the energy region of interest, so this stability result is satisfactory.

\subsubsection{Device stability}
\label{subsubsec:GKrS:system-stability:slow-control-stability}
The magnetic field, temperature, and pressure were examined in different sections of the experiment. In the WGTS the relative stability (i.e., the ratio of the standard deviation to the mean of the data set) of the magnetic field was better than \num{2e-5} on an hourly basis, based on the current readouts of the superconducting magnets. The hourly temperature stability was \num{1e-4}. The relative stability values over the whole GKrS campaign of 7 days are \num{2e-5} for the magnets and \num{1.2e-3} for the temperature, demonstrating the long-term stability of WGTS operation.

The stabilities of the CPS and pre-spectrometer magnets were analyzed in the same way, resulting in relative stabilities of better than \num{7e-4} over the entire run. The pressure in the main spectrometer showed a relative stability of \num{7.4e-2}, reflecting the very low mean pressure in the denominator; the baffle temperature, essential for background reduction, was stable to within \num{1.7e-3}. Stability values were comparable during the CKrS measurement phase (\sref{sec:GKrS-campaign}).

The apparatus operated within specifications during these measurement periods, and we do not expect device instability to significantly affect count rates.

\subsubsection{High-voltage ripple}
\label{subsec:GKrS:high-voltage-ripple}
During these measurements, the high-voltage system was used without the post-regulation system, as described in \sref{par:high-voltage}. 
The ripple-probe of the post-regulation was read out with an oscilloscope, showing a \SI{50}{\hertz} sine wave with an amplitude of \SI{0.19}{\volt} at \SI{17}{\kilo\volt} and \SI{0.21}{\volt} at \SI{30}{\kilo\volt}, respectively~\cite{Rest2018}. These values were subsequently used for spectrum analysis.

The ripple could also be estimated from FPD count-rate fluctuations, providing an independent cross-check for the oscilloscope estimation. However, the sensitivity was limited by lack of statistics and by fluctuations of the grid cycles. A dedicated ripple measurement with a new cycle recorder is planned for future measurement campaigns.

\subsubsection{Monitor-spectrometer data}
\label{subsubsec:GKrS:system-stability:monitor-spectrometer-data}
\begin{figure}[tbp]
	\centering
	\includegraphics[width=0.8\textwidth]{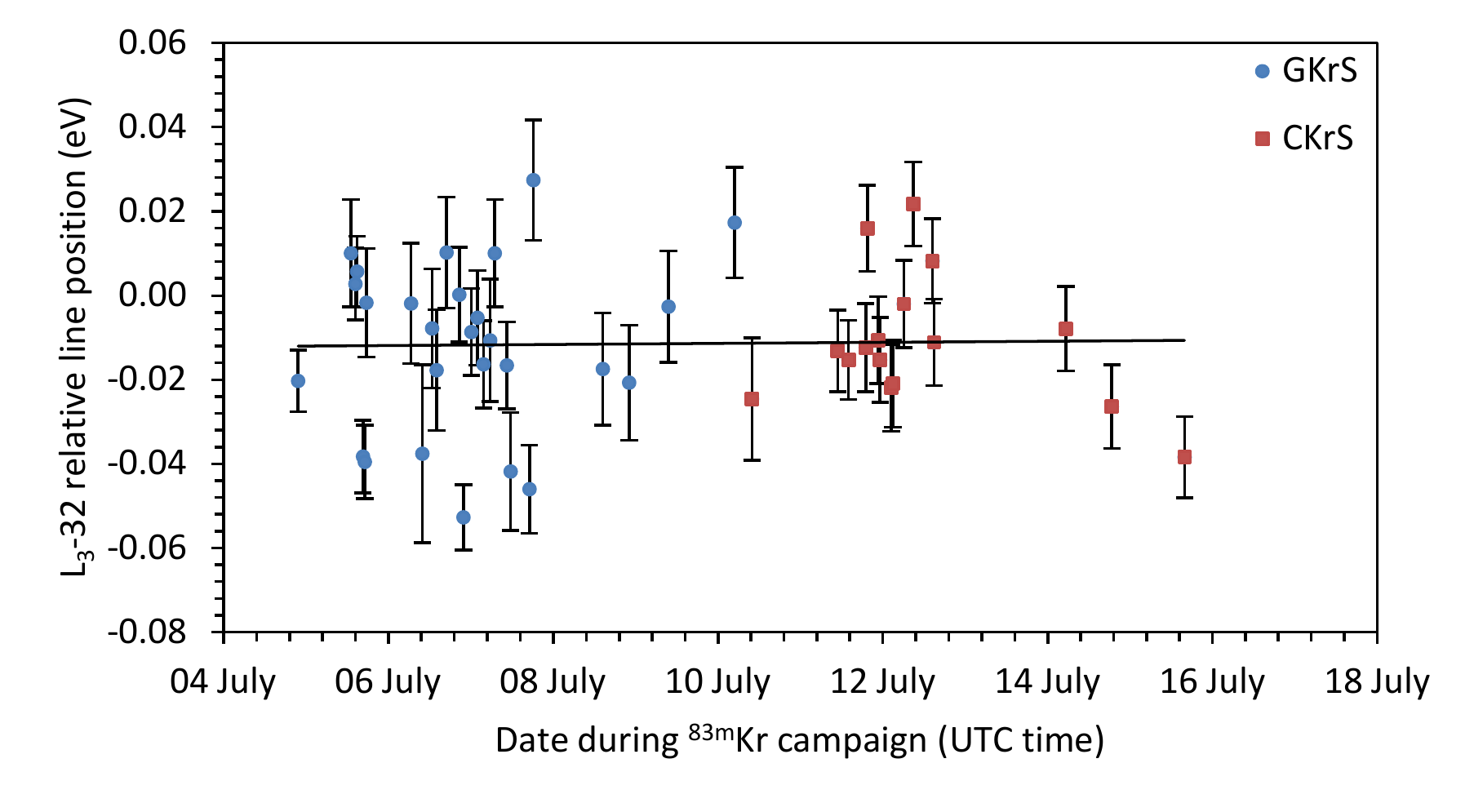}
	\caption{The L$_3$-32 relative line position measured at the monitor spectrometer with the implanted \textsuperscript{83}Rb/\kr{} HOPG source during the GKrS and CKrS campaigns.}
	\label{fig:L-32_relative_line_position}
\end{figure}

The stability of the L$_3$-32 line position was also investigated at the monitor spectrometer (\sref{sec:mos}) to verify the stability of the high-voltage system. This line was periodically measured during line scans and reference runs within the \SIrange[range-phrase=--]{30465}{30480}{\volt} range, between measurements of other \kr{} lines using the main spectrometer. For the entire measurement campaign with both sources, a single implanted \textsuperscript{83}Rb/\kr{} source with an \SI{8}{\kilo\electronvolt} implantation energy was used. The line was fit with a Doniach-Sunjic line shape~\cite{Doniach1970} convolved with a Gaussian function and with the calculated spectrometer transmission function. The fit of the L$_{3}$-32 integral spectra was performed with the \textsc{MoSFitter} software package~\cite{Slezak2013}. The variation in line position in these spectra fell into the interval of $\pm$\SI{50}{\milli\electronvolt} for each of the GKrS and CKrS campaigns, as shown in \fref{fig:L-32_relative_line_position}. The variation meets the KATRIN design requirements~\cite{KATRIN2005}.

The line positions in \fref{fig:L-32_relative_line_position} were fit with a straight line by the least-squares method. As a result the mean line-position drift was determined to be \SI{14+-15.6}{\milli\electronvolt\per month} for the GKrS and CKrS campaigns combined; the measurement period of only one week for each source type was too short to obtain a more precise value.

\section{Condensed \kr{} campaign}
\label{sec:CKrS-campaign}

As described in \sref{sec:CKrS}, a dedicated condensed krypton source (CKrS) will be available for routine calibration of high-voltage properties during neutrino-mass operations. This source was tested and characterized \textit{in situ} in July~2017, just after the gaseous krypton measurement campaign described in \sref{sec:GKrS-campaign}. Since the source is mounted on the CPS, upstream portions of the beamline (the rear section, the WGTS, and the DPS) were separated by closed valves and shut down. The CPS beam tube was operated at \SI{4.6}{\kelvin}, maintaining a beamline pressure of order \SI{e-9}{\milli\bar}. The spectrometers and focal-plane detector were operated at \SI{e-11}{\milli\bar}; the pre-spectrometer was maintained at ground potential. The monitor spectrometer was operated as a reference.

\secref{ssec:ckrs_first_electrons} describes first light from the CKrS in the FPD as well as demonstrating source motion. Qualitative information about the line shapes is given in \sref{ssec:ckrs_qualitative_information_about_line_shapes}. Rate stability is reported in \sref{ssec:ckrs_stability}.

\subsection{First electrons}
\label{ssec:ckrs_first_electrons}
During the CKrS campaign several radioactive krypton films were prepared by first cleaning the substrate via heating and laser ablation before freezing on the krypton. Then the valve to the \textsuperscript{83}Rb generator was left open for continously condensing fresh \textsuperscript{83m}Kr to achieve a stable count rate. This sequence was repeated for a total of three times. For each film, energy scans of the emitted electrons were performed and the filtered electrons were registered at the FPD.

\begin{figure}
	\centering
	\includegraphics[width=0.45\linewidth]{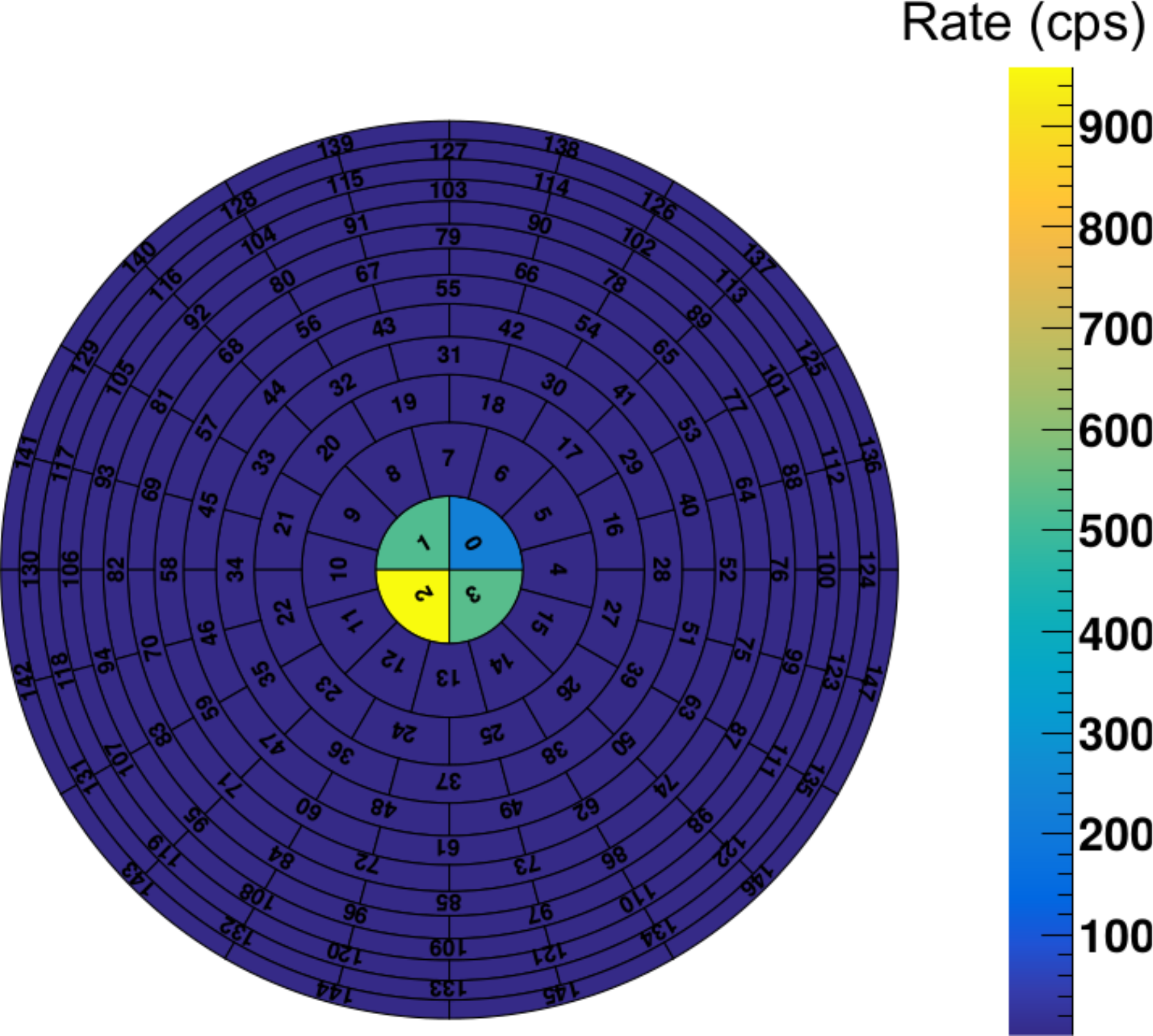}
	\includegraphics[width=0.45\linewidth]{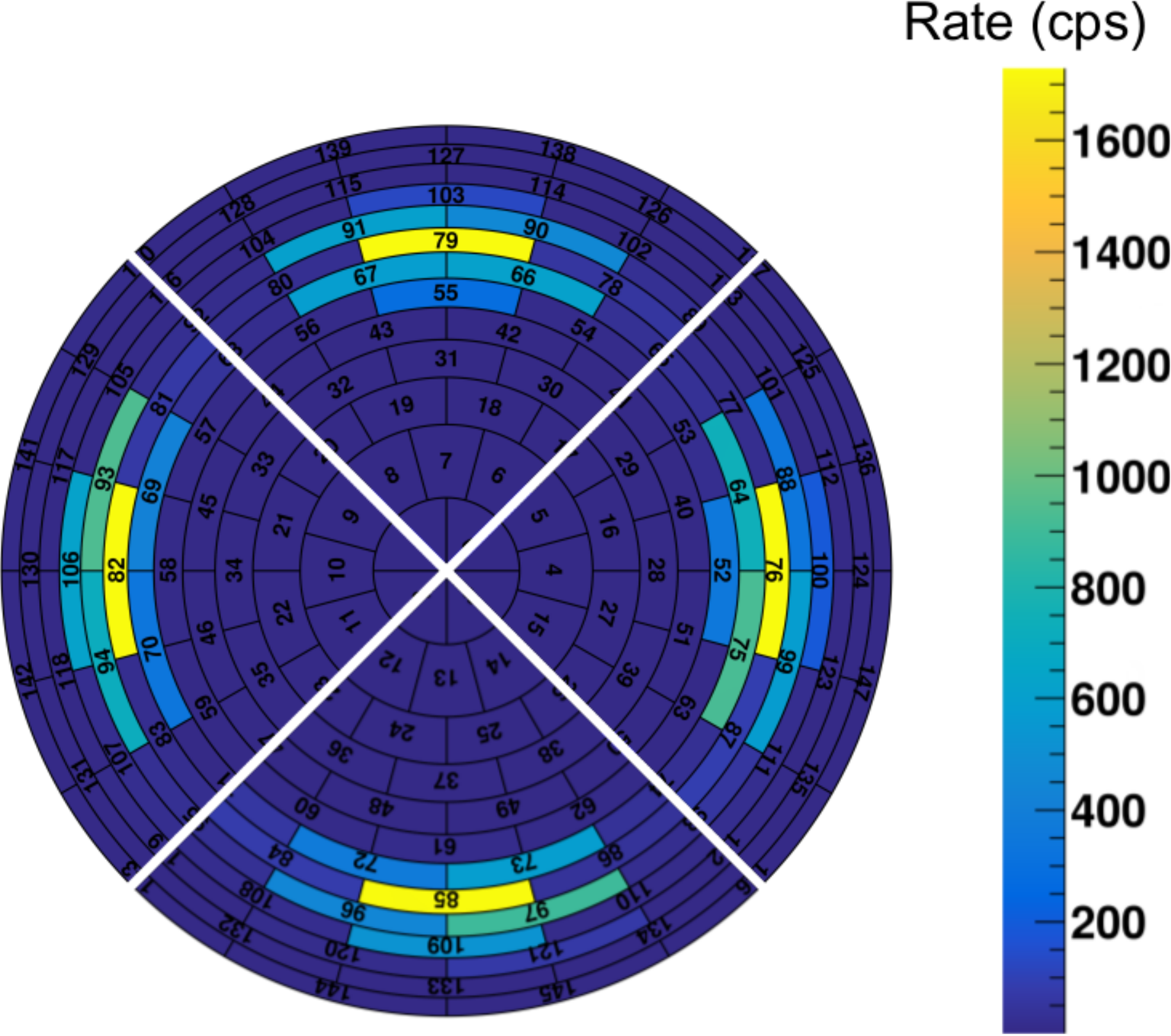}
	\caption{FPD pixel map of electron events originating from the CKrS. \textbf{Left:} Hit pattern produced by the CKrS after it was aligned with the center of the beamline for a scan of the M$_2$-32 and M$_3$-32 lines. \textbf{Right:} A mosaic showing K-32 line scans at four different CKrS positions resulting in different areas of the detector being illuminated. The color scale represents the hit rate on the FPD (cps).}
	\label{fig:ckrs_fpd_plot_and_ckrs_movement}
\end{figure}

For the very first measurements with the CKrS, the substrate was inserted into the beamline and the valve to the \textsuperscript{83}Rb generator was opened to prepare the first film. Radioactive \kr{} moved slowly through the capillary towards the \SI{26}{\kelvin} substrate where it condensed and eventually decayed. The operation of the CKrS elevated the measured background to about \SI{20}{cps} due to decays of uncondensed \kr{} in the CPS and spectrometer volumes. The CKrS itself produced a much higher rate, with about \SI{2000}{cps} initially observed in the FPD even with the main spectrometer set to a high retarding voltage of \SI{31.2}{\kilo\volt} as a detector-safety precaution. The electron rate increased as an increasing amount of krypton was frozen onto the substrate, and saturated over time (\sref{ssec:ckrs_stability}). Using the FPD, the source could be aligned so that it was located in the center of the flux tube, as shown in \fref{fig:ckrs_fpd_plot_and_ckrs_movement} (left). After this basic proof of principle, the main spectrometer was used to scan the lines of the \kr{} spectrum.

The construction of the CKrS permits vertical and horizontal positioning of the active source area anywhere within the cross section of the magnetic flux tube (\sref{sec:CKrS}). During operation, the CKrS substrate region was positioned in the center of the beam tube as a starting position. At the beginning of the CKrS measurement campaign the positioning system was tested and proved fully operational. \figref{fig:ckrs_fpd_plot_and_ckrs_movement} (right) displays four different positions of the substrate with a condensed \kr{} film as it illuminated the FPD. In each case the hit pattern indicated a clear center of events, with the distribution of pixels reflecting the extent (\SI{2}{\centi\metre} in diameter) of the substrate.

\subsection{Line shapes}
\label{ssec:ckrs_qualitative_information_about_line_shapes}
The full high-resolution \kr{} conversion electron spectrum was measured during the CKrS campaign. The condensed source showed a much higher rate per illuminated pixel than the GKrS, but only a few pixels were illuminated at a time. With the CKrS centrally positioned, and once the rate had stabilized, the spectral lines were scanned with \SI{0.5}{\volt} steps upwards and downwards. Over the course of the measurement campaign the three different films were used to scan all spectral lines of the \num{9.4} (L, M, and N) and \SI{32}{\kilo\electronvolt} (K, L, M, and N) transitions. Most lines were scanned with an analyzing plane magnetic field of \SI{0.27}{\milli\tesla}. With this magnetic-field setting, the main-spectrometer energy resolution (as given by  \eref{eq:mac-e}) was $\Delta E = $ \SI{2.06}{\electronvolt} for \SI{32}{\kilo\electronvolt} electrons. For the narrow N lines, the voltage steps were decreased to \SI{0.1}{\volt}, and a lower magnetic field of \SI{0.1}{\milli\tesla} was used to improve to the energy resolution to $\Delta E = $ \SI{0.8}{\electronvolt} for \SI{32}{\kilo\electronvolt} electrons.

\begin{figure}[tbp]
	\centering
	\includegraphics[width=.8\textwidth]{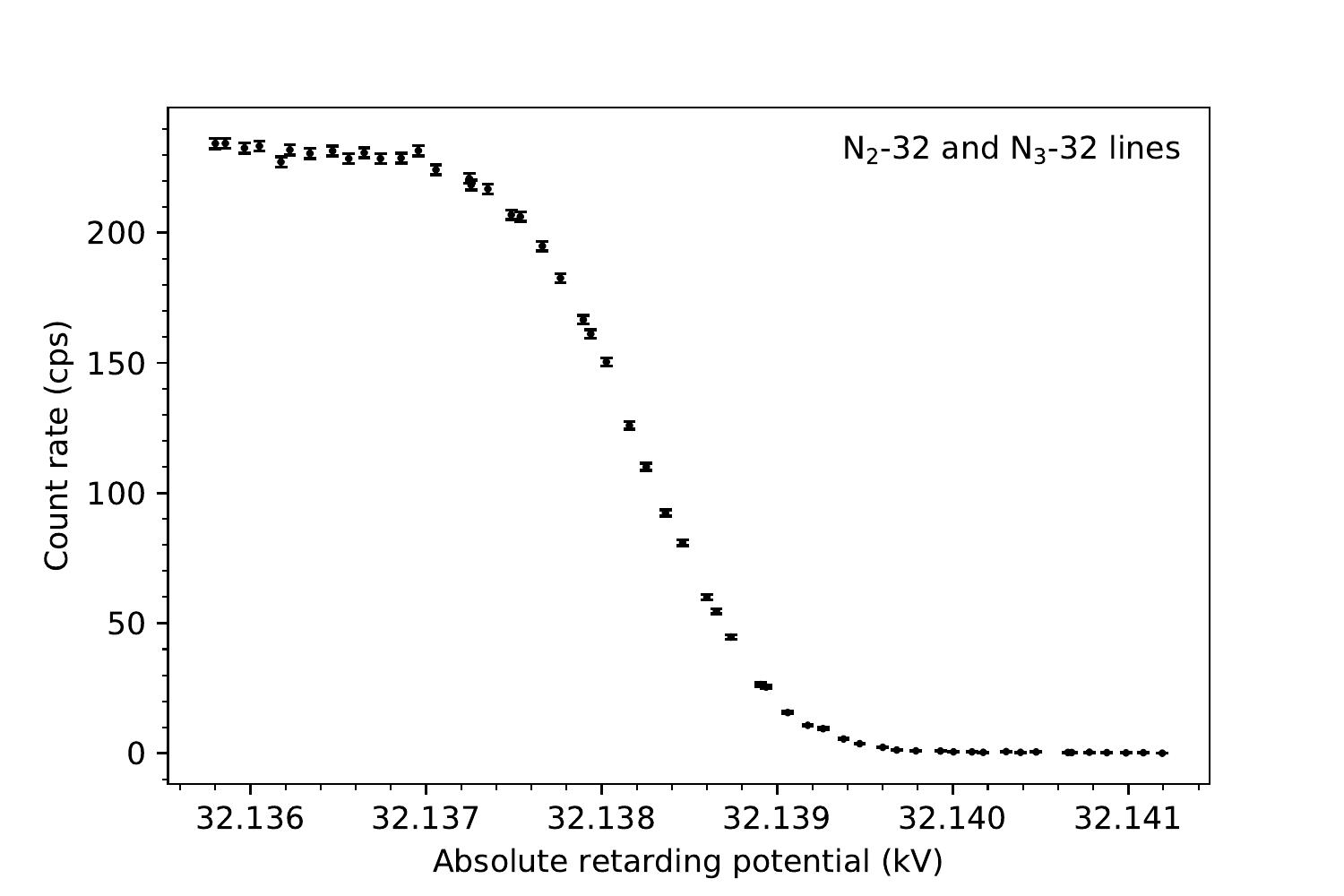}
	\caption{Rate trend of the N$_2$ and N$_3$ line doublet of the \SI{32}{\kilo\electronvolt} transition measured at the FPD, versus the absolute value of the retarding potential in the analyzing plane. Since the CKrS illuminated a small portion of the whole detector wafer, only events from the four central pixels are included. The energy values were corrected to reflect the mean potential at the central pixels (\sref{apparatus:sds}) while other corrections (e.g., work-function differences between source substrate and main spectrometer) are left to a forthcoming detailed analysis.}
	\label{fig:ckrs_n_line}
\end{figure}

\figref{fig:ckrs_n_line} shows the integrated measurement of the N$_2$ and N$_3$ doublet with sub-\SI{}{\electronvolt} resolution. This demonstrates that the CKrS could provide a sufficient rate of conversion electrons even for lines with low intensity, as the combined branching ratio to these two lines is about \SI{0.75}{\percent}~\cite{venos:2018}.

\subsection{CKrS system stability}
\label{ssec:ckrs_stability}
\begin{figure}[tbp]
	\centering
	\includegraphics[width=.8\textwidth]{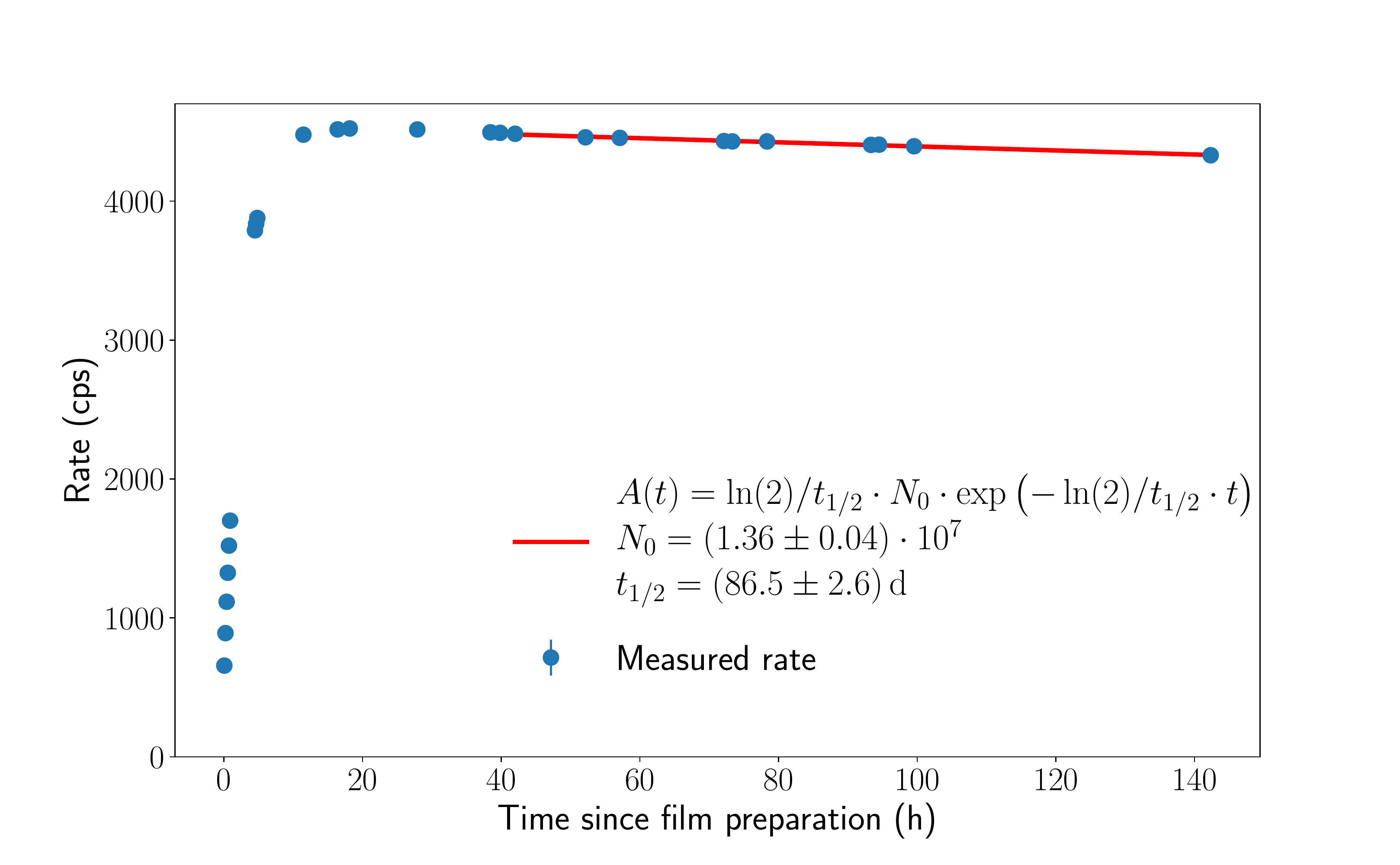}
	\caption{Change of the activity of the CKrS after film preparation, as measured with the FPD. The earliest data points show that it took about \SI{15}{\hour} for the rate to stabilize after a new film began to condense onto the substrate. The later data points were fitted with an exponential decay for the activity, as shown by the line.}
	\label{fig:ckrs_rate_trend}
\end{figure}

In addition to the normal line scans, special runs were taken during the campaign to investigate the time dependence of the conversion-electron rate. In contrast to spectroscopic measurements, the retarding potential of the main spectrometer was kept at a fixed value of \SI{31.21}{\kilo\electronvolt} during these stability runs, each with a run length of 5--10~min. After a new film was prepared, we found that it took approximately \SI{15}{hours} for the rate to stabilize, as shown in \fref{fig:ckrs_rate_trend}. By design, during calibration measurements using the CKrS, the source is situated in the beamline and blocks electrons from the WGTS. Hence, neutrino-mass measurements cannot be performed at the same time. As a result of the findings described here, new \kr{} films will be condensed outside of the flux tube about \SI{15}{hours} before their use for calibration, thus minimizing measurement downtime.

We also studied the behavior of the activity over longer time scales, taking stability runs over more than \SI{140}{hours}. As shown in \fref{fig:ckrs_rate_trend}, the rate trend $A(t)$ after stabilization followed the decay of the \textsuperscript{83}Rb generator over time $t$:
\begin{equation}
	A(t) = \lambda \cdot N_0 \cdot \exp\left(-\lambda\cdot t\right),
\end{equation}

\noindent where $N_0$ represents the initial activity after stabilization. The best-fit half-life of $t_{1/2} = \ln 2 / \lambda =$ \SI{86.5 \pm 2.6}{\day} is in agreement with the literature value of \SI{86.2 \pm 0.1}{\day}~\cite{Mccutchan2015}, within statistical uncertainties.
Only the data points after 40 hours were used in the fit, to ensure an equilibrium state between the \kr{} gas flow through the capillary and the decay rate on the substrate.

\section{Discussion and conclusion}
\label{sec:conclusion}

Precision spectroscopy of tritium \betadec{} in the kinematic endpoint region is a sensitive and model-independent means of probing the absolute neutrino-mass scale. The currently most mature technique, based on a strong gaseous molecular tritium source and a large electrostatic spectrometer of MAC-E filter type, is used by the \ke{} with the goal of improving the present sensitivity on $m_{\bar{\nu}_\mathrm{e},\, \mathrm{eff}}$ by an order of magnitude to \SI{0.2}{\electronvolt} (\SI{90}{\percent} CL). KATRIN is currently preparing for the start of a five-year measurement phase, in which it will collect the necessary statistics to reach its sensitivity goal.

With the arrival of the last major system components on site by autumn 2015, the complex KATRIN experimental facility has been completed after an extensive construction and assembly period. Integration of the full beamline, comprising source and transport units as well as the spectrometer and detector section, was accomplished in 2016. While accounts of the characterization of individual components have been given elsewhere (for instance in~\cite{Grohmann2013,Fraenkle2017,Behrens2017}), in this work we report on first commissioning results using the entire 70-meter beamline of the experiment. The corresponding measurements were performed in two major operational runs. In autumn 2016, the First Light campaign utilized \SI{100}{\electronvolt} photoelectrons from the rear section to establish obstruction-free electron transport without adiabaticity losses over the full distance on to the main detector. In the same measurement phase, a source of non-radioactive ions was used to mimic the creation of positive ions in the tritium source, allowing thorough testing of ion transport, detection and removal concepts. Building on the First Light results, we conducted further system characterization measurements during a two-week campaign in July 2017 with mono\-energetic conversion electrons from the decay of \kr{}. 

The measurements were carried out using three different and complementary realizations of krypton sources: an implanted \kr{} source attached to the monitor spectrometer (equipping a secondary beamline which is connected to the same retarding potential as the main spectrometer), gaseous \kr{} injected into the beam tube of the windowless gaseous source cryostat, and finally a condensed \kr{} source inserted at the source-side entrance of the spectrometer section. The data presented here comprise about five days of measurement each with the gaseous and condensed \kr{} sources, and regular parallel runs taken with the implanted source at the monitor spectrometer throughout the entire period. This mode of data-taking not only allowed testing, for the first time, of the operation of the gaseous and condensed krypton sources as well as of the front and rear beam monitor instrumentation, but it also exercised the transition between operational modes as will be required during routine running of KATRIN. 

It is worth emphasizing the dual purpose of the krypton measurement campaign: While serving as a commissioning run for the various krypton sources to be routinely used for systematics control during neutrino-mass runs, it likewise allowed us to exploit the unique properties of the mono\-energetic and isotropically emitted conversion electrons from a nuclear standard to characterize the precision MAC-E filter as well as overall system alignment and stability. The wide span of conversion electron lines from \kr{} between \SI{7}{\kilo\electronvolt} and \SI{32}{\kilo\electronvolt}, at natural line widths on the order of just a few \SI{}{\electronvolt}, provides an ideal tool for such instrumental tests. The results of these analyses are reported in this article. Moreover, these measurements enabled us to develop and test a novel calibration method for the absolute energy scale~\cite{Rest2018}. First spectroscopic analyses of the conversion electron lines will be reported in~\cite{Slezak2018}.

We find that the MAC-E filter of the KATRIN main spectrometer works as a precision instrument with sharp energy resolution and excellent transmission characteristics, fulfilling the demanding requirements for the neutrino mass scan of the tritium \betaspec{} around \SI{18.6}{\kilo\electronvolt}. These findings are also an important basis for KATRIN's extended physics program which encompasses, for example, the search for potential signatures of light sterile neutrinos a few eV away from the spectral endpoint (see, e.g.,~\cite{SejersenRiis2011a,Formaggio2011,Esmaili2012}), and the TRISTAN detector as a future extension of the KATRIN setup to look for keV-scale sterile neutrinos in the high-rate part deep down in the \betadec{} spectrum~\cite{Mertens2015,Adhikari2016}. 

All systems involved in the measurement chain -- from the temperature stabilization of the source beam tube and the precision high-voltage feed for the retarding potential, to the focal-plane detector and data-acquisition systems -- demonstrate stability and performance well within (and often even surpassing) specifications. These accomplishments provide the foundation for the start of tritium data-taking with KATRIN.

\acknowledgments
We acknowledge the support of Helmholtz Association (HGF), 
Ministry for Education and Research BMBF (05A17PM3, 05A17PX3, 05A17VK2, and 05A17WO3), 
Helmholtz Alliance for Astroparticle Physics (HAP), and Helmholtz Young Investigator Group (VH-NG-1055) in Germany; 
Ministry of Education, Youth and Sport (CANAM-LM2011019, LTT18021), in cooperation with JINR Dubna (3+3 grants) in the Czech Republic;
and the Department of Energy through grants DE-FG02-97ER41020, DE-FG02-94ER40818, DE-SC0004036, DE-FG02-97ER41033, DE-FG02-97ER41041, DE-AC02-05CH11231, and DE-SC0011091 in the United States.

\printnomenclature





\providecommand{\href}[2]{#2}\begingroup\raggedright\endgroup

\end{document}